\documentclass[a4paper,10pt]{article}

\usepackage{contribution}
\usepackage{graphicx}
\usepackage{lineno}

\newcommand{\weblink}[2][]{%
    \ifthenelse{\equal{#1}{}}%
    {\textnormal{\url{#2}}}%
    {\textnormal{\href{#2}{#1}}}%
}

\def\beq{\begin{equation}}
\def\eeq#1{\label{#1}\end{equation}}
\def\eeqn{\end{equation}}

\def\beqa{\begin{eqnarray}}
\def\eeqa#1{\label{#1}\end{eqnarray}}
\def\eeqan{\end{eqnarray}}

\let\bar=\overbar

\def\Dslash{\not{\hbox{\kern-4pt $D$}}}
\def\dslash{\not{\hbox{\kern-2pt $\del$}}}

\def\msb{{\bar{\ssstyle M \kern -1pt S}}}

\newcommand{\contribution}[7][]{%
  \clearpage
  \thispagestyle{plain}

  \ifthenelse{\equal{#1}{}}
  {\hypersetup{pdftitle={#2}}}
  {\hypersetup{pdftitle={#1}}}
  \hypersetup{pdfauthor={{#3} {#4}}}
  {\centering\normalfont\LARGE\bfseries\sffamily #2 \par\nobreak}
  \lhead{}
  \chead{%
    \textit{\footnotesize XXIInd International Workshop ``High-Energy Physics and Quantum Field Theory'', 
June 24 -- July 1, 2015, Samara, Russia}%
  }
  \rhead{}
  \bigskip
  \begin{center}
    {#3} {#4}\ifthenelse{\equal{#6}{}}{}{\footnote{\weblink[#6]{mailto:#6}}}
    \ifthenelse{\equal{#7}{}}{}{#7} \\
    \textit{#5}
  \end{center}
  \bigskip
}

\renewcommand{\abstract}[1]{%
  \begin{center}
    \begin{minipage}{0.85\textwidth}
      \begin{footnotesize}
        #1
      \end{footnotesize}
    \end{minipage}
  \end{center}
  \bigskip
}

\begin{document} 

%
%
%
%
%
%
{  


%



\contribution[Higgs boson discovery and recent results]  
{Higgs boson discovery and recent results}  
{Martin}{Flechl}  
{Institute of High Energy Physics,\\
Austrian Academy of Sciences,\\
Nikolsdorfergasse 18, 1050 Vienna, Austria}
{martin.flechl@cern.ch}  
{on behalf of the CMS collaboration}  
%

\abstract{%
After briefly discussing the discovery of a Higgs boson at the Large Hadron Collider, 
an overview of recent results in Higgs boson physics obtained with the CMS experiment is presented. 
The focus is on measurements of the properties of the recently discovered Higgs boson with a mass of about 125 GeV. A brief selection of results 
in searches for Higgs bosons beyond the standard model is given, and prospects of future Higgs boson measurements and searches are discussed.
}
%





\section{Introduction}
In July 2012, the ATLAS~\cite{atlas} and CMS~\cite{cms} collaborations announced the discovery~\cite{disc_atlas,disc_cms} of a new particle compatible with 
being the Higgs boson of the standard 
model (SM). Subsequent measurements of the properties of this particle are all consistent with the SM Higgs boson interpretation. 
In the following sections, the most recent results from studies of this Higgs boson using the data collected with the CMS experiment are presented. 
First, the development of the discovery is discussed briefly. Then, the most recent results for various Higgs boson production and decay channels are given, 
followed by the measurement of Higgs boson properties using a combination of these results. 
Searches for Higgs bosons predicted by theories beyond the standard model (BSM) are briefly summarized and prospects for future Higgs boson property 
measurements are discussed. Most of the results are based on the full data set recorded in 2011 and 2012 at center-of-mass energies of 7 TeV and 8 TeV, 
corresponding to an integrated luminosity of about 5 and 20 fb$^{-1}$, respectively.

\section{Higgs boson discovery at the LHC}
Precise predictions from the theory community (e.g. the LHC Higgs Cross Section Working Group), 
excellent tools due to previous experiments (e.g. parton distribution functions and HERA) and restrictions on the 
Higgs boson phase space (e.g. on its mass, due to LEP and the Tevatron) have facilitated 
the Higgs boson discovery and subsequent measurements of the Higgs boson properties.
Certainly, the discovery would have been impossible without the excellent operation 
of the LHC by the accelerator team and the world-wide LHC Computing Grid. However, in the following the focus will be entirely on the 
experimental results.

Before the first LHC results, the Higgs boson mass had been constrained by the LEP experiments to be larger than 114.4 GeV~\cite{leplimit} and by the
Tevatron experiments to be either smaller than 156 GeV or larger than 177 GeV~\cite{tevlimit}. The LHC searches thus focused on the region above 114 GeV. 
At the EPS conference in July 2011, the first LHC exclusion of SM Higgs mass values was presented: 
The CMS experiment excluded most of the range from 145 GeV to 400 GeV, albeit with two small gaps from 216 GeV to 226 GeV and from 288 GeV to 310 GeV. 
Incidentally, the CMS experiment also observed an excess of about 3 Gaussian standard deviations ($\sigma$) with respect to the hypothesis of an SM  
without a Higgs boson, at the $m_H$ values of 118 GeV, 144 GeV and 162 GeV~\cite{eps2011}. The quoted significance here corresponds to a local $p$-value.

At HCP in November 2011, the first combined ATLAS and CMS results were presented~\cite{hcp2011}. This allowed the exclusion 
of the SM Higgs boson over a large mass range, from 141 GeV to 476 GeV. Together with indirect constraints such as from electroweak 
fits, this essentially narrowed down the likely region for the SM Higgs boson, if it exists, to a narrow mass range from 
114 GeV to 141 GeV. At the same time, the data set showed two excesses with about 3 $\sigma$ (local) at the Higgs boson masses of 119 GeV and 145 GeV. 
However, this corresponds to a global significance of only about 1.6 $\sigma$.

At a CERN seminar in December 2011, the CMS results allowed the exclusion of Higgs boson mass values from 
127 GeV to 600 GeV~\cite{cern2011}. The only surviving excess of events was observed at $m_H=125$ GeV albeit with a 
global significance of only 0.8 $\sigma$. However, it was very intriguing that the results shown by ATLAS 
at the same seminar~\cite{atlascern2011} gave essentially the same picture.

Finally, at another CERN seminar in July 2012, both ATLAS and CMS announced the observation of a new boson, 
compatible with SM Higgs bosons predictions~\cite{cern2012,atlascern2012}. Both experiments observed an excess at about $m_H=125$ GeV each with a local significance of 
about 5 $\sigma$. This day also marks the transition to the Higgs boson measurement era: In the same seminar, measurements of the Higgs boson mass (with an 
uncertainty of about 1 GeV) and of the Higgs boson signal strength for various channels were presented, all in good agreement with the SM Higgs expectations.

\section{Higgs boson search results}\label{sec:chan}
\begin{figure}[htb]
\begin{center}
\begin{minipage}{0.46\textwidth}
\includegraphics[height=6.0cm]{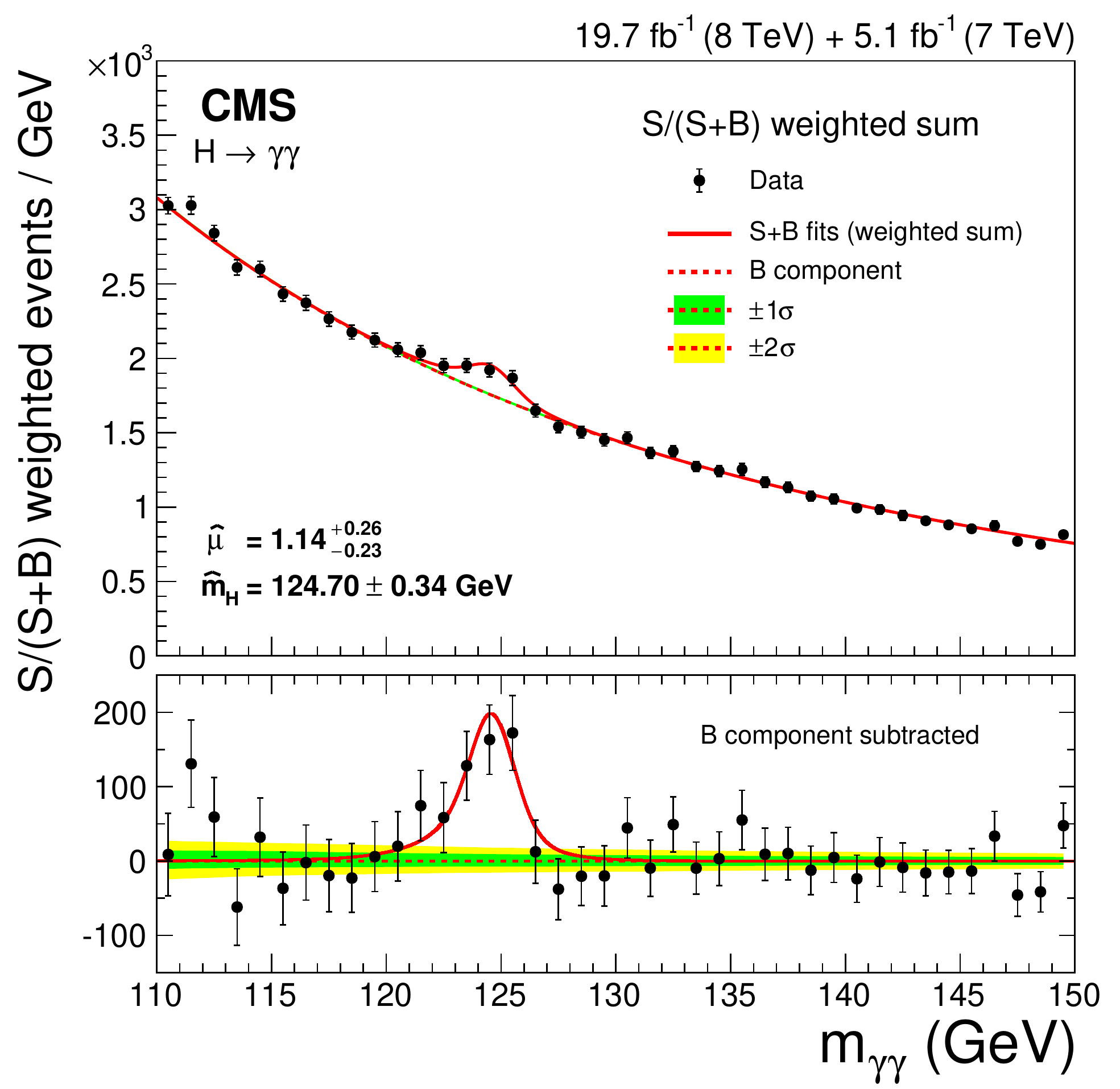}
\caption{\label{hgg_c1}The $m_{\gamma\gamma}$ distribution as weighted sum of all categories~\cite{hgg_cms}. $S$ and 
$B$ are the number of signal and background events, respectively.} 
\end{minipage}\hspace{0.05\textwidth}%
\begin{minipage}{0.46\textwidth}
\includegraphics[height=6.0cm]{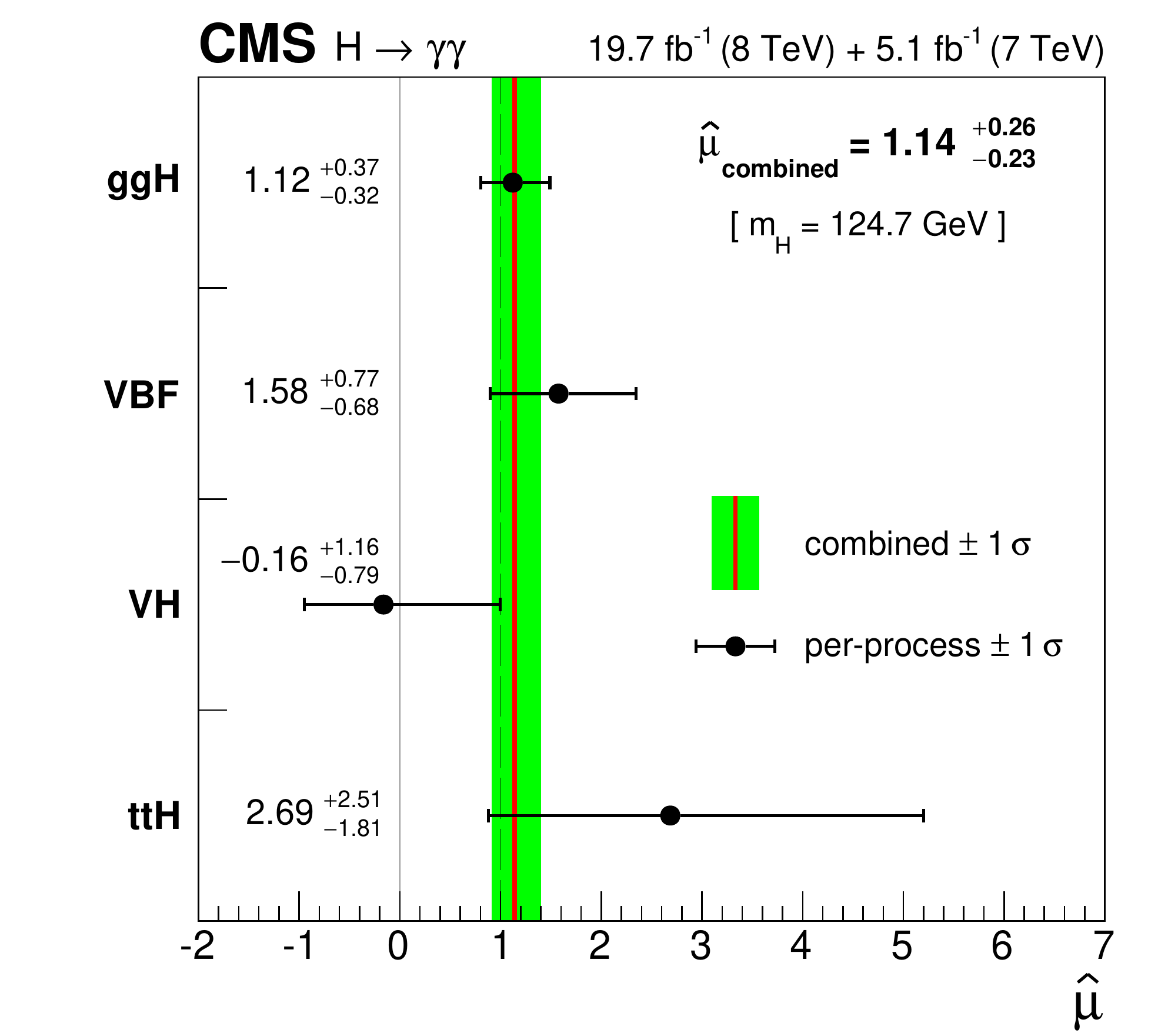}
\caption{\label{hgg_a1}Signal strength in the $H \to \gamma\gamma$ channel. Shown is the result for the individual Higgs production modes 
and their combination~\cite{hgg_cms}. }
\end{minipage}
\end{center}
\end{figure}
The dominant Higgs boson production modes are gluon-gluon fusion (ggF), 
vector boson fusion (VBF), production in association with a vector boson V ($VH$, $V=W$ or $V=Z$), 
and in association with top quarks ($ttH$). 
The five most sensitive Higgs boson decay channels at the LHC are 
the modes $\gamma\gamma$, 4-leptons, $WW$, $\tau\tau$ 
and $b\bar{b}$~\cite{yr1,yr2,yr3}. Results for these channels are 
summarized in the following.

\subsection{$H \to \gamma\gamma$}
The $H \to \gamma\gamma$ channel is characterized by relatively high total event counts but a low signal-over-background ratio. 
Furthermore, it offers a high mass resolution ($m_{\gamma\gamma}$) and thus contributes strongly to the Higgs boson mass 
measurement. The analysis~\cite{hgg_cms} proceeds by selecting events with two photons in various categories, aimed at different production 
modes and regions with high signal-to-background ratio. 
This is done by additionally requiring two jets with a high rapidity gap (VBF), additional leptons and in some cases 
missing energy ($VH$) or an event topology consistent with a $t\bar{t}$ event ($ttH$). The final discriminating variable 
is the $m_{\gamma\gamma}$ estimator. The background, dominantly continuum $\gamma\gamma$, $\gamma$+jet and di-jet events, is estimated by 
fitting the sidebands of the $m_{\gamma\gamma}$ distribution, as illustrated in Fig.~\ref{hgg_c1}. The signal is visible on top of the estimated 
background at $m_{\gamma\gamma} \approx 125$~GeV.

Slightly more $H \to \gamma\gamma$ candidate events than expected are observed; however, the measurement is within 
one Gaussian standard deviation ($\sigma$) of the SM expectation. The signal strength $\mu$ (observed cross section times 
branching ratio divided by the SM expectation) is also measured individually for the various Higgs boson production modes, see Fig.~\ref{hgg_a1}. 
All values are in agreement with the SM expectation of $\mu=1$. The observed signal significance is $5.7\sigma$ (expected: $5.2\sigma$).

\subsection{$H \to 4l$}
\begin{figure}[htb]
\begin{center}
\begin{minipage}{0.43\textwidth}
\includegraphics[height=6.5cm]{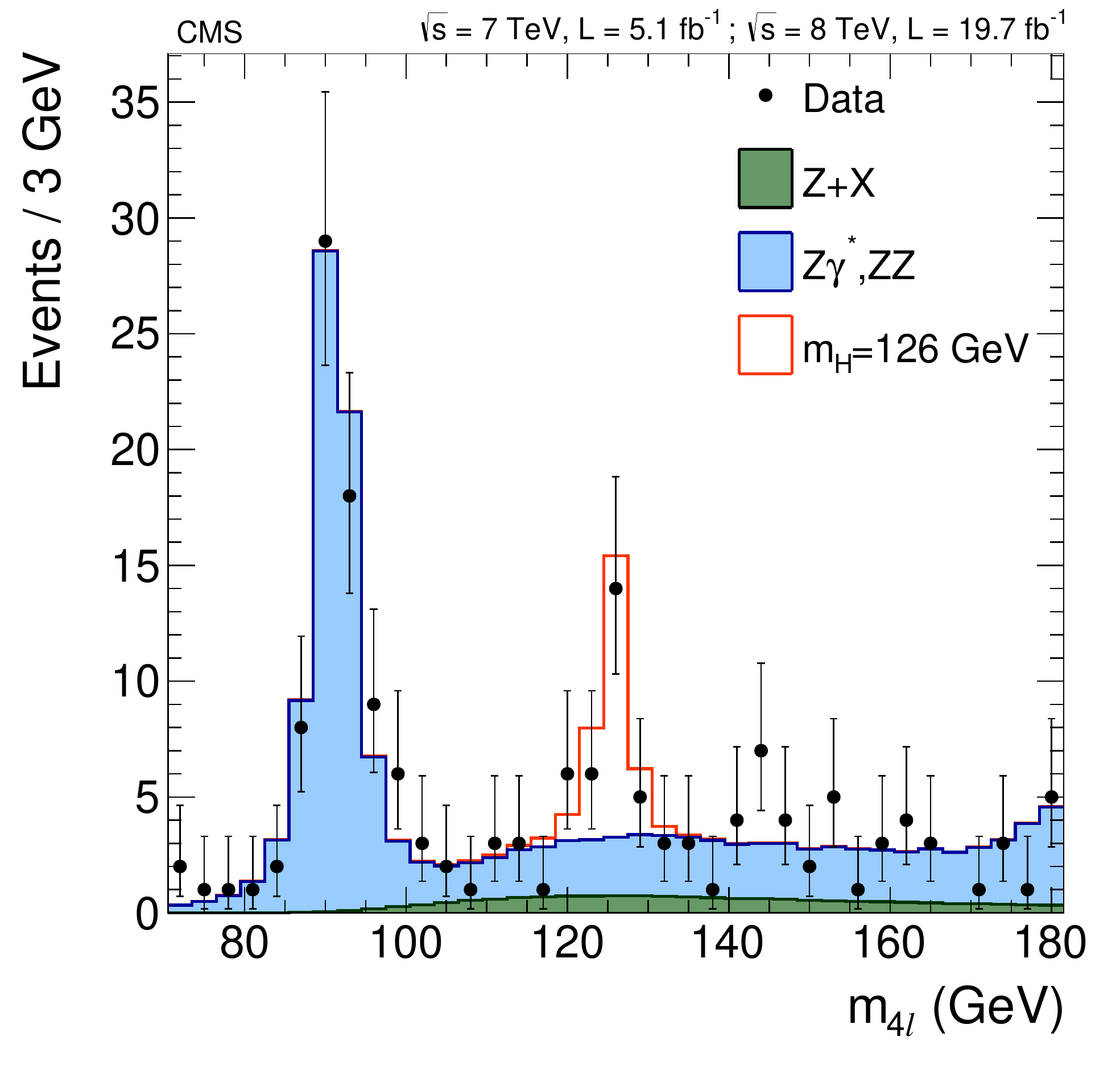}
\caption{\label{h4l_a1}Combined $m_\mathrm{4l}$ distribution~\cite{h4l_cms}.} 
\vspace{1\baselineskip}
\end{minipage}\hspace{0.04\textwidth}%
\begin{minipage}{0.43\textwidth}
\includegraphics[height=7cm]{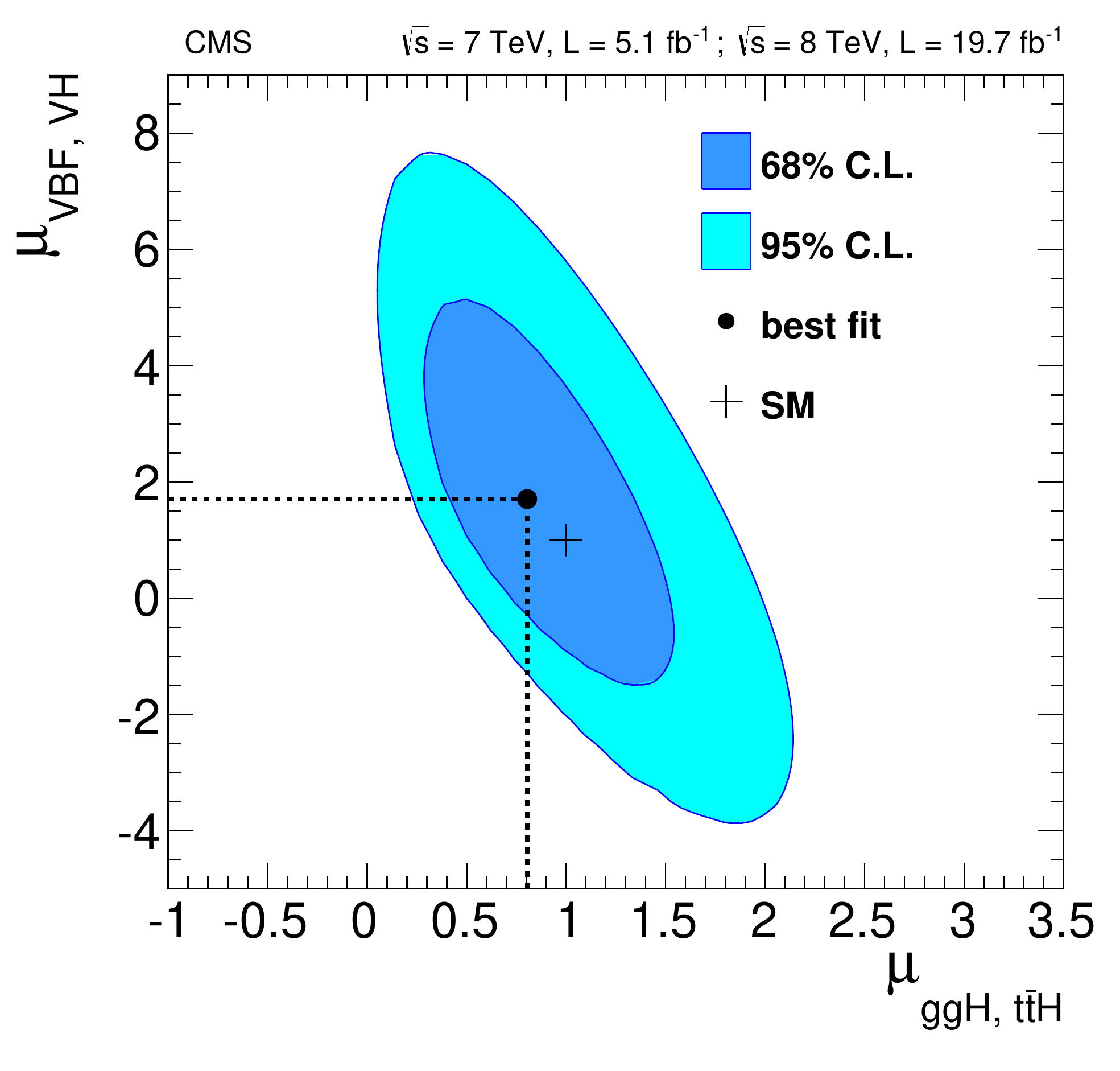}
\caption{\label{h4l_c1}Measurement of the signal strength associated to ggF and $ttH$  production 
versus VBF and $VH$ production~\cite{h4l_cms}.}
\end{minipage}
\end{center}
\end{figure}
The expected rate of $H \to 4$ lepton events is low compared to the other channels presented here; however, this is compensated by the highest 
signal-over-background ratio. This channel is further characterized by a high mass resolution ($m_\mathrm{4l}$) and dominates the Higgs boson mass 
measurement together with $H \to \gamma\gamma$. The analysis~\cite{h4l_cms} first requires four light leptons and is then pursued in 
three subchannels, $4e$, $4\mu$ and $2e2\mu$. Like the $H \to \gamma\gamma$ analysis, it is split into categories aimed at different production modes. 
However, the main sensitivity rests in the $0/1$ jet category which is mostly sensitive to ggF production. The main background is continuum $ZZ^*$ production 
which is estimated from simulation and validated in control regions. The combined $m_\mathrm{4l}$ distribution is shown in Fig.~\ref{h4l_a1}. 
The mass peak at about 125 GeV is clearly visible.

The results are in very good agreement with the SM expectation. This is illustrated in Fig.~\ref{h4l_c1} where the signal strength 
for the ggF and $ttH$ production mode is measured with respect to the VBF and $VH$ production modes. 
The observed signal significance is $6.8\sigma$ (expected: $6.7\sigma$). 

\subsection{$H \to WW$}
\begin{figure}[htb]
\begin{center}
\begin{minipage}{0.47\textwidth}
\includegraphics[height=6.9cm]{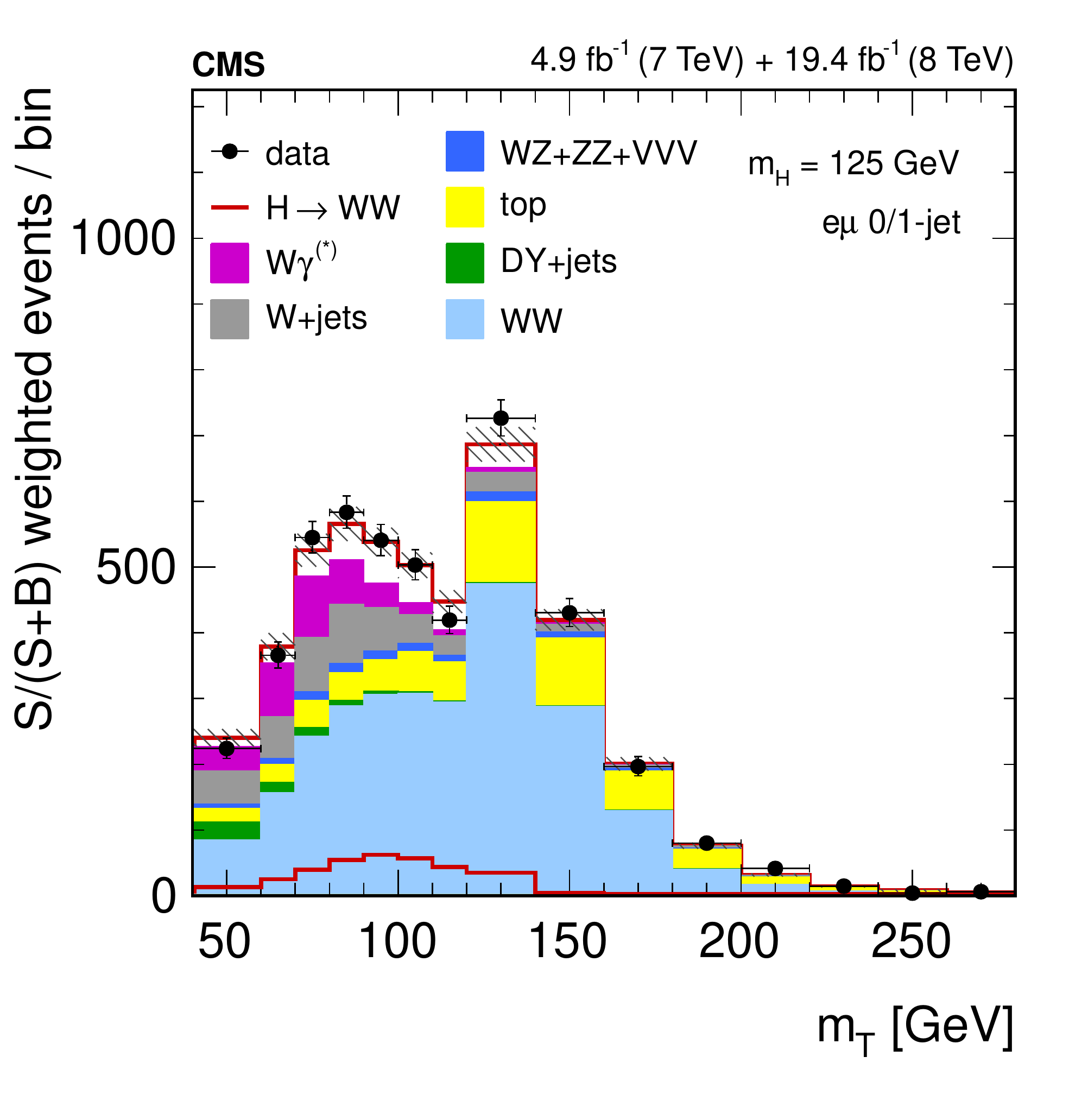}
\caption{\label{hww_a1}The $m_T$ distribution in the category ``$e \mu$ + 0/1 jet''~\cite{hww_cms}. 
Events are weighted by $S/(S+B)$ of the corresponding bin of the final analysis discriminant, where 
$S$ and $B$ are the number of expected signal and background events, respectively.}   
\end{minipage}\hspace{0.04\textwidth}%
\begin{minipage}{0.47\textwidth}
\includegraphics[height=6.9cm]{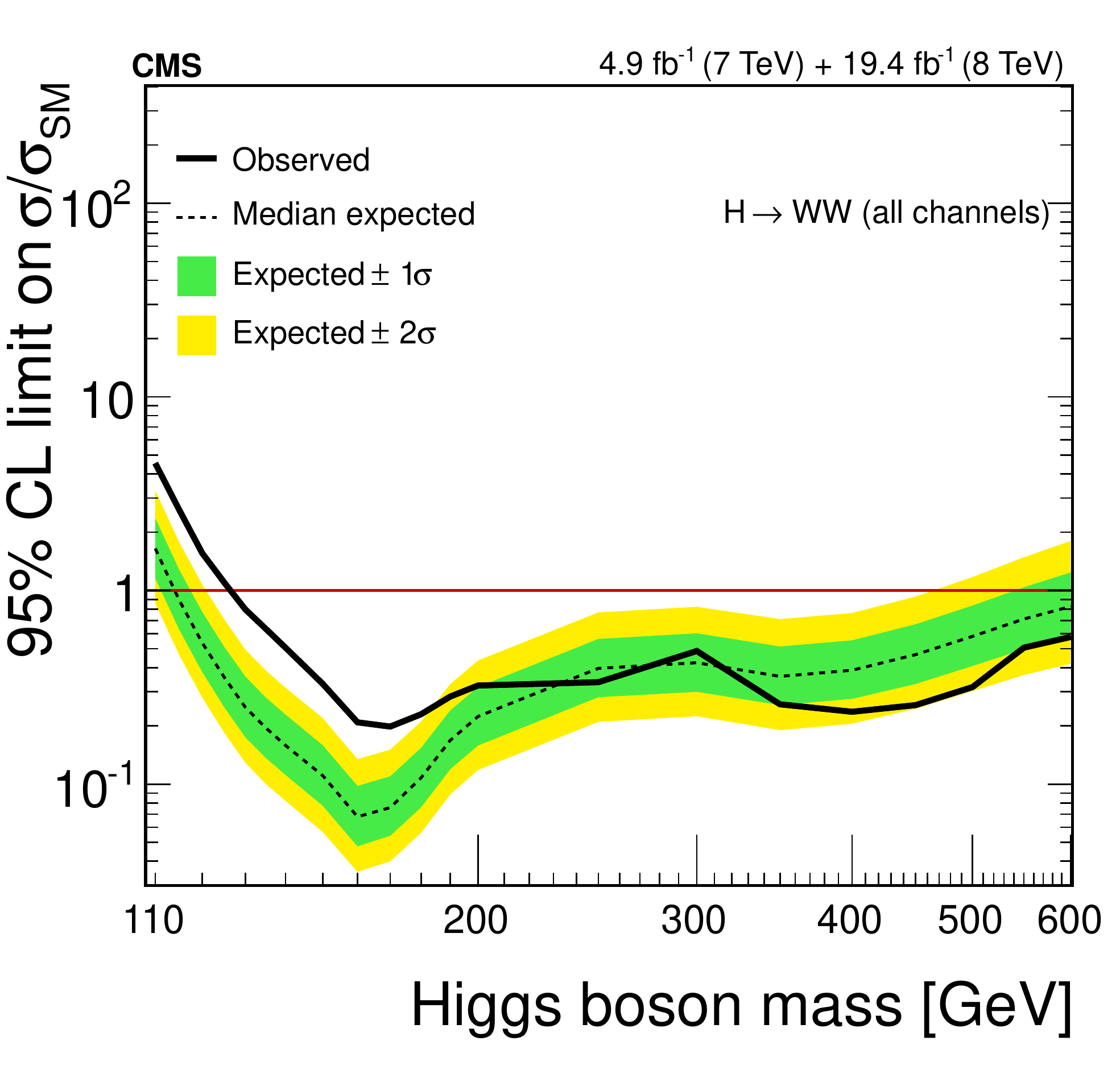}
\caption{\label{hww_c1}The excluded region as function of the Higgs boson mass in the $WW$ channel~\cite{hww_cms}.}
\vspace{3\baselineskip}
\end{minipage}
\end{center}
\end{figure}
The branching ratio for Higgs boson decays to a $W$ boson pair (0.22) is the second-highest after the $b\bar{b}$ mode. It thus offers 
a relatively high expected rate but suffers from a high irreducible $WW$ continuum background which is hard to suppress due to 
the low $m_{WW}$ resolution, caused by the presence of neutrinos in the sensitive $W$ decay modes. The $W$+jets and top quark pair 
background is also sizable. The most sensitive subchannel features a $e\nu \mu\nu$ final state with low jet 
multiplicity~\cite{hww_cms}. Requirements of additional jets or leptons aim to add extra sensitivity by exploiting 
also the VBF, $VH$ and $ttH$ production modes. The $m_T$ distribution in the category ``$e\nu \mu\nu$ + 0/1 jet'' is shown in 
Fig.~\ref{hww_a1}.

The observed rate is within about $1\sigma$ of the SM expectation. In Fig.~\ref{hww_c1}, the Higgs boson 
exclusion versus $m_H$ is shown. Visible are both the excess of events at about 125~GeV and the exclusion of additional heavy SM-like Higgs bosons.

\subsection{$H \to \tau\tau$}
\begin{figure}[htb]
\begin{center}
\begin{minipage}{0.47\textwidth}
\includegraphics[height=7cm]{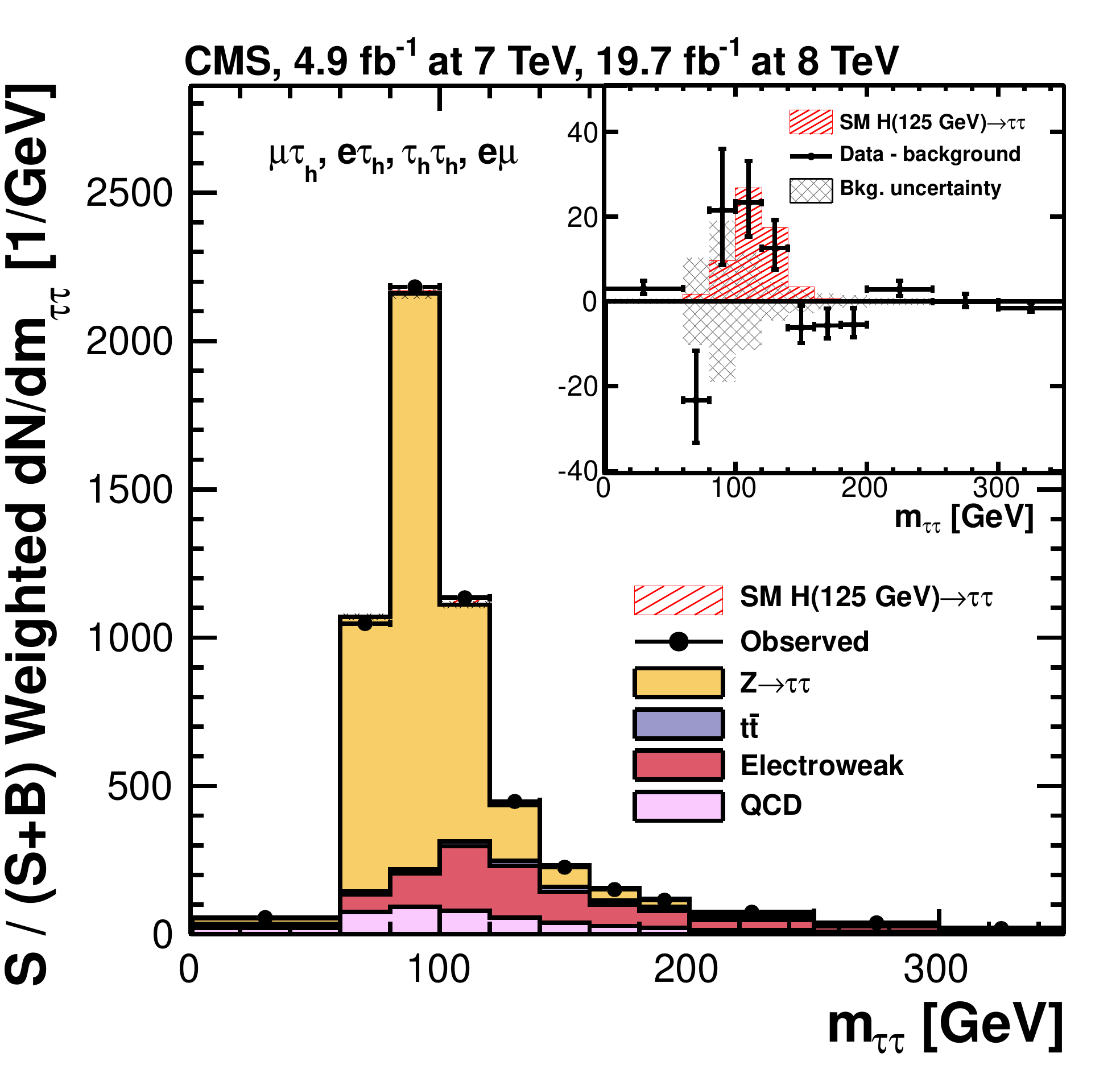}
\caption{\label{htt_c1}The $m_{\tau\tau}$ distribution~\cite{htt_cms}. Events of several channels are weighted by signal purity and then combined. 
The insert at the upper right shows the background-subtracted distribution.} 
\end{minipage}\hspace{0.04\textwidth}%
\begin{minipage}{0.47\textwidth}
\includegraphics[height=7.8cm]{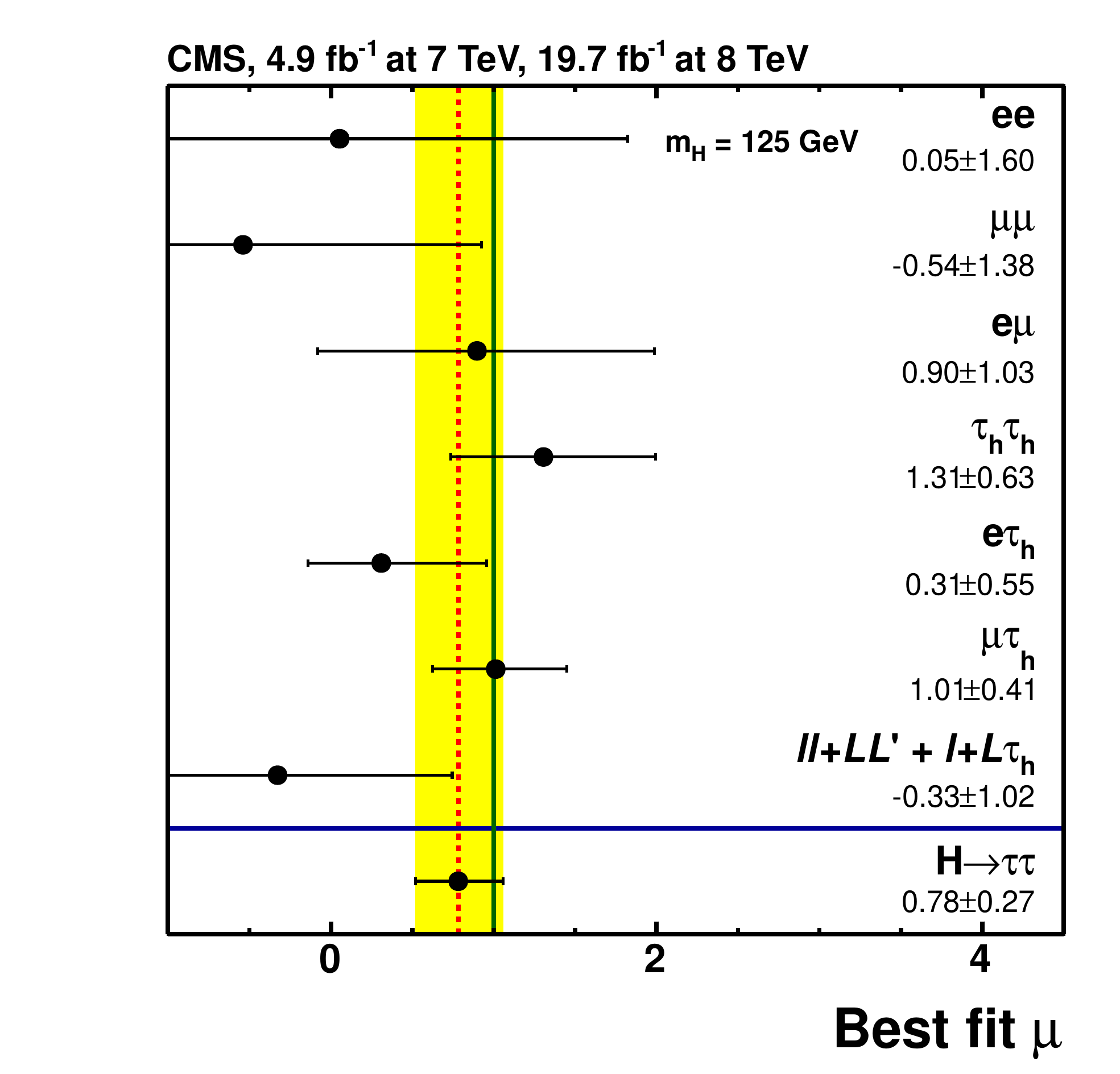}
\caption{\label{htt_a1}The measured signal strength for various $H \to \tau\tau$ final states~\cite{htt_cms}. }
\end{minipage}
\end{center}
\end{figure}
%
The $H \to \tau\tau$ analysis is one of the most complex analyses of LHC data: Six different channels, depending on 
the combination of $\tau$ lepton decays, are investigated: $ee$, $e\mu$, $\mu\mu$, $e\tau_h$, $\mu\tau_h$ and $\tau_h\tau_h$ (here, $\tau_h$ signifies 
a hadronic $\tau$ lepton decay, and neutrinos are omitted). For all these channels, categories motivated by the production mode can be 
established: 0/1-jet, boosted (boson candidate with high $p_T$), VBF and $VH$~\cite{htt_cms} as well as $ttH$~\cite{tth_cms}. 
The analysis is based on orthogonal selection requirements and implements almost one hundred categories. 
An $m_{\tau\tau}$ estimator using kinematic information of the whole decay chain to reconstruct a mass value in spite of the presence of 
a number of undetected neutrinos is used. An example is shown in Fig.~\ref{htt_c1}. The main background are $Z/\gamma^* \to \tau\tau$ events which 
are estimated by replacing muons in $Z \to \mu\mu$ collision data events with simulated $\tau$ leptons.

$H \to \tau\tau$ decays are observed with a significance of $3.2\sigma$ (expected: $3.7\sigma$). Together with the corresponding ATLAS result~\cite{htt_atlas}, this 
constitutes the first evidence for Higgs boson decays to leptons. The signal strength measured individually for different final states are all in agreement with the 
SM expectation, see Fig.~\ref{htt_a1}.

\subsection{$H \to bb$}
\begin{figure}[htb]
\begin{center}
\begin{minipage}{0.47\textwidth}
\includegraphics[height=7.0cm]{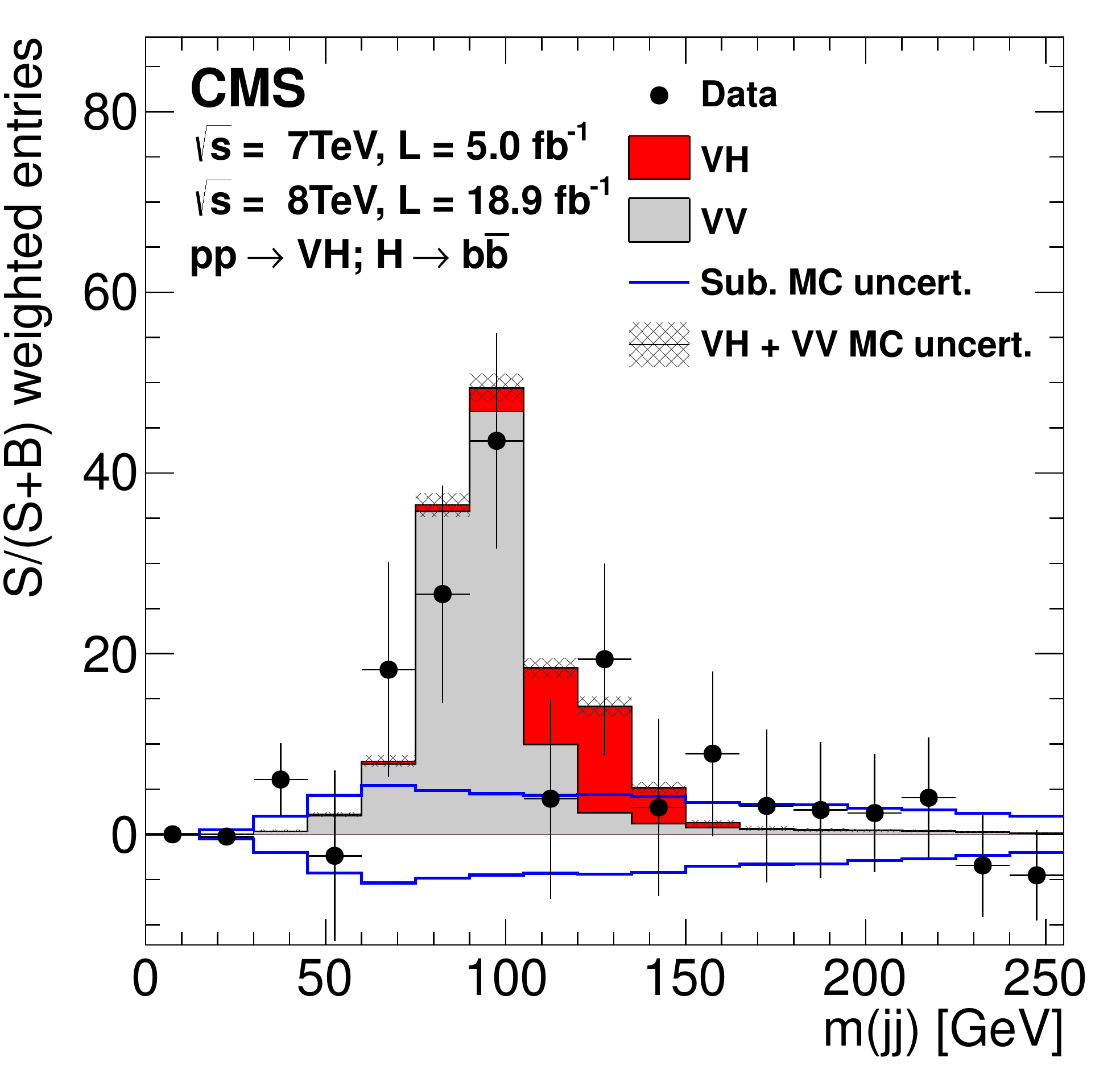}
\caption{\label{hbb_a1}The $m_{bb}$ distribution after subtracting all backgrounds except diboson events for the $VH$ analysis~\cite{hbb_cms}.} 
\end{minipage}\hspace{0.04\textwidth}%
\begin{minipage}{0.47\textwidth}
\includegraphics[height=7.0cm]{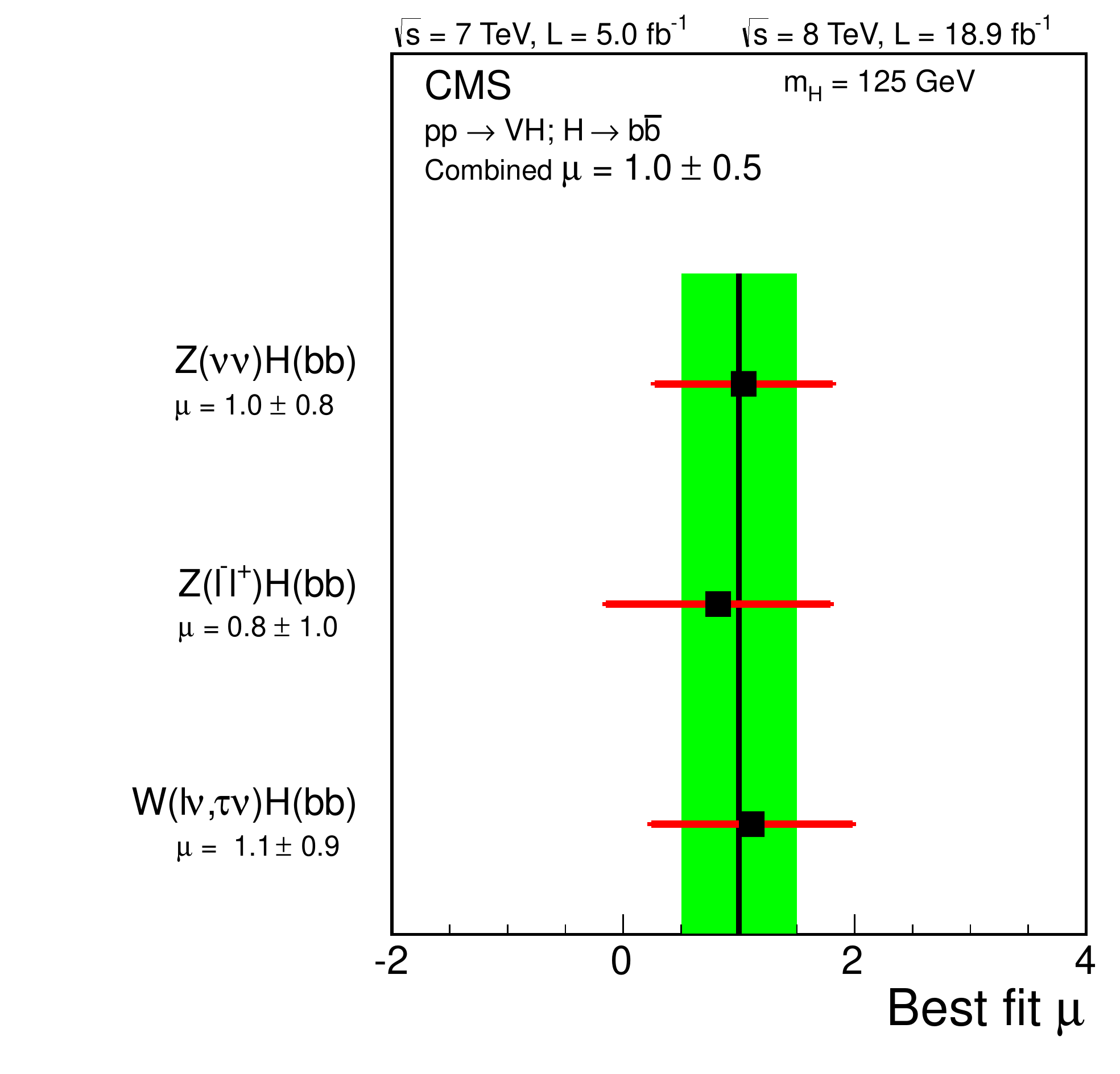}
\caption{\label{hbb_c1}The signal strength in the $H \to bb$ channel as measured in the $VH$ analysis~\cite{hbb_cms}.}
\end{minipage}
\end{center}
\end{figure}
The $H \to bb$ analysis is extremely challenging at a hadron collider, despite the high expected branching ratio of 
$0.58$ at $m_H=125$ GeV. The ggF production is not accessible as the signal cannot be separated from the 
overwhelming non-resonant $b\bar{b}$ background.
The situation for VBF production is slightly better. The main workhorse, however, are $VH$ events; in addition, 
$ttH$ is also investigated. 
The $VH$ analysis~\cite{hbb_cms} uses BDTs to isolate the signal 
in an overwhelming background of $Z$, $W$, $t\bar{t}$, $VV$ and multi-jet events. The $m_{bb}$ distribution after 
background subtraction is shown in Fig.~\ref{hbb_a1} and 
the measured signal strength in Fig.~\ref{hbb_c1}. Both observed and expected signal significance are $2.1\sigma$.

The VBF analysis~\cite{hbb_vbf_cms} uses dedicated trigger items and a BDT to categorize the signal candidate events. 
Then, the $m_{bb}$ distribution is fit for each category. A statistically insignificant excess of events is observed, 
leading to an observed signal significance of $2.2\sigma$ where only $0.8\sigma$ are expected.
The combined significance for the $VH$, VBF and $ttH$ analysis is $2.6\sigma$ (expected: $2.7\sigma$).

\subsection{$ttH$}
\begin{figure}[htb]
\begin{center}
\begin{minipage}{0.57\textwidth}
\includegraphics[height=6.3cm]{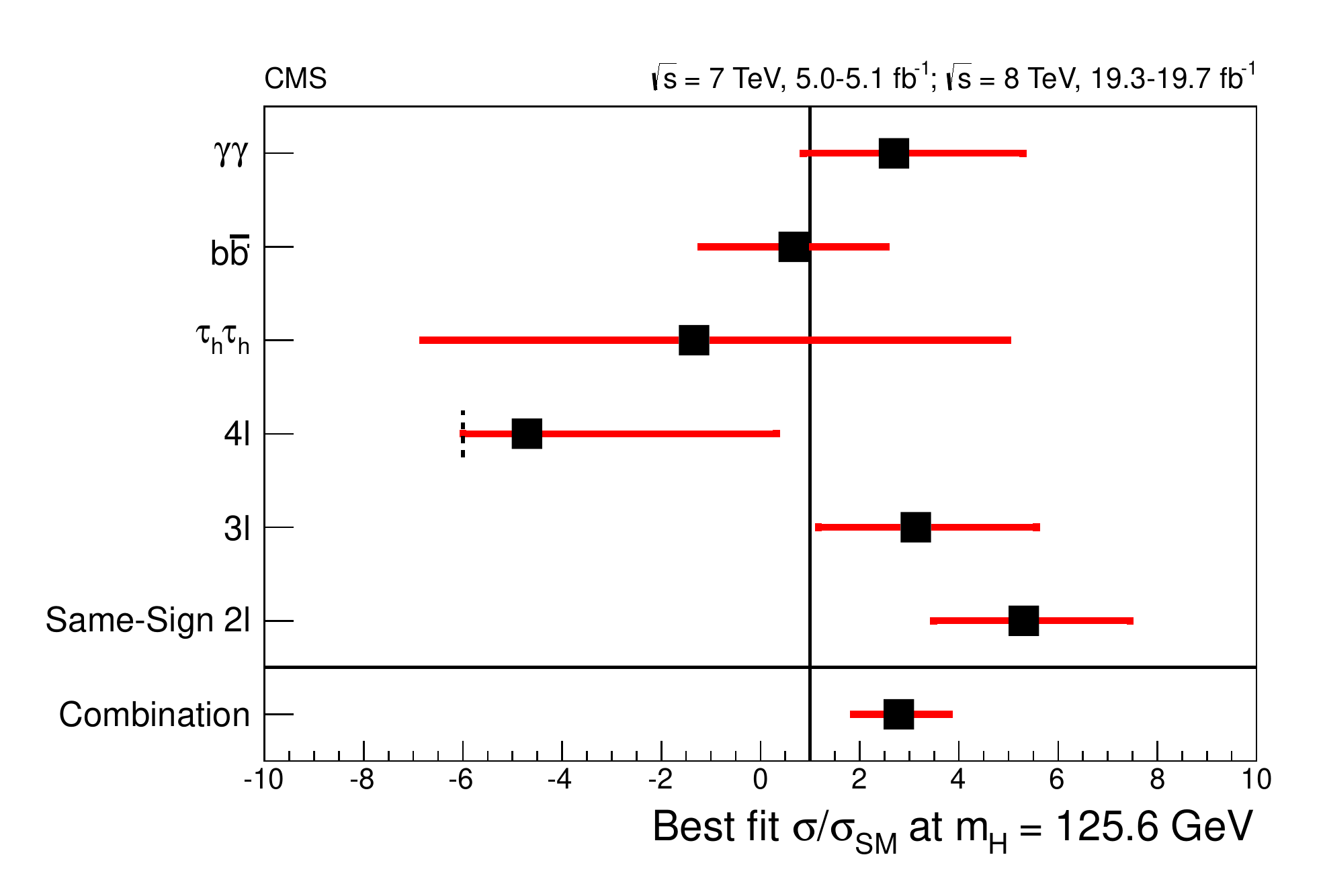}
\caption{\label{tth_c2}The measured signal strength in the $ttH$ analyses~\cite{tth_cms}.}
\vspace{1\baselineskip}
\end{minipage}\hspace{0.04\textwidth}%
\begin{minipage}{0.37\textwidth}
\includegraphics[height=6.3cm]{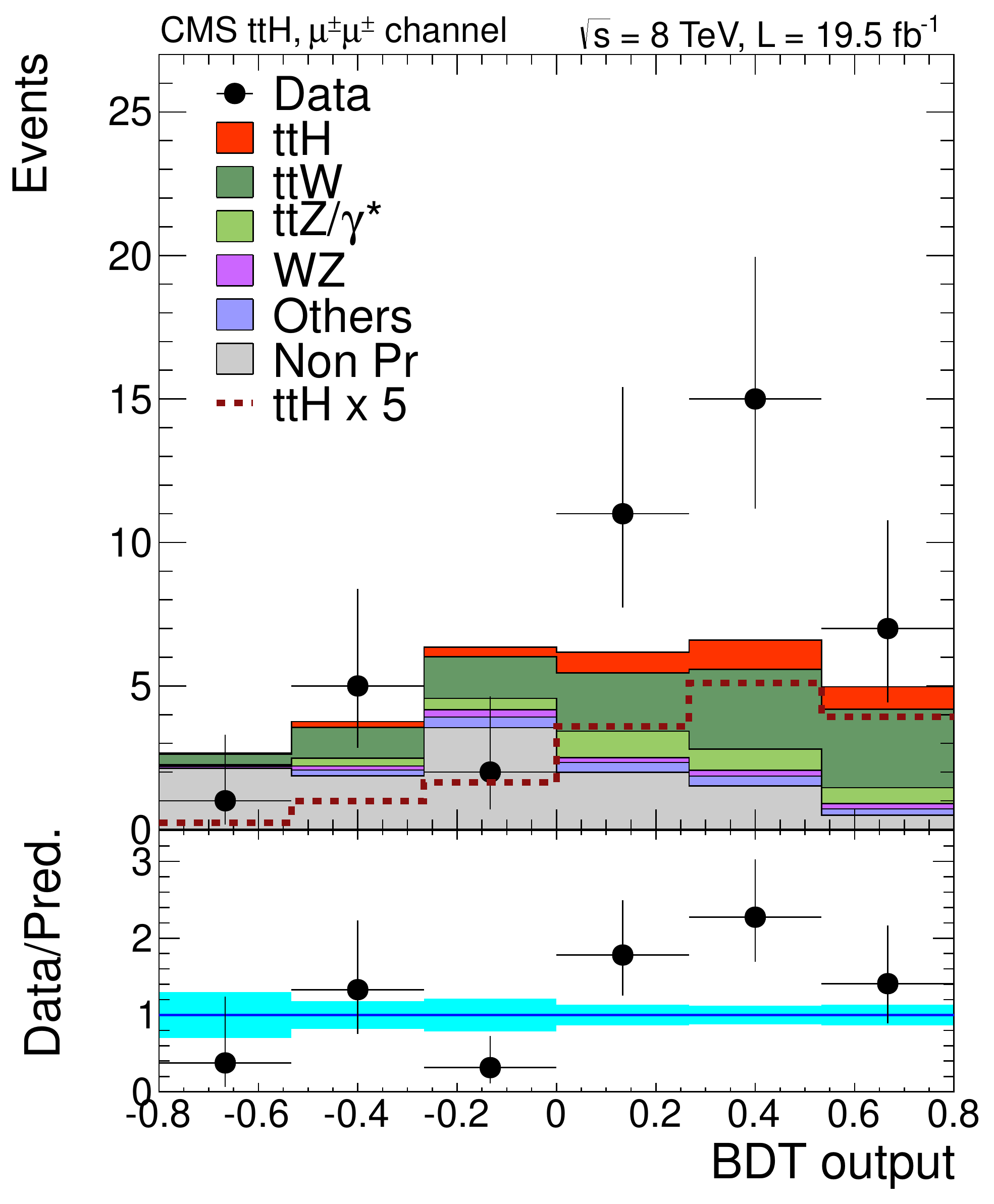}
\caption{\label{tth_c1}The BDT output for the $ttH$ analysis in the same-sign two-lepton channel~\cite{tth_cms}.} 
\end{minipage}
\end{center}
\end{figure}
The $ttH$ channel poses very particular challenges and is thus in most cases not treated in the context of the usual decay mode analyses. 
Instead, dedicated $ttH$ analyses and combinations are produced~\cite{tth_cms}. Six different final states are investigated, see 
Fig.~\ref{tth_c2}. The combined signal strength measured is $\mu=2.8 \pm 1.0$ and thus about $2\sigma$ high compared to the SM expectation. 
The excess of events is almost entirely driven by the same-sign two-lepton category. The BDT output for this category is shown in Fig.~\ref{tth_c1} 
and illustrates this point.

\section{Higgs boson property measurements}
The Higgs boson property measurements use the results of the analyses aiming at the various production and decay modes presented in the previous 
section. For each measurement, the channels sensitive to the property in question are combined. Among the studied properties are mass, width, 
signal strength (i.e., normalized cross section times branching ratio), coupling strength and tensor coupling structure (CP properties) of the 
Higgs boson in relation to other SM particles.

\subsection{Mass}
\begin{figure}[!h]
\begin{center}
\begin{minipage}{0.43\textwidth}
\includegraphics[height=6.0cm]{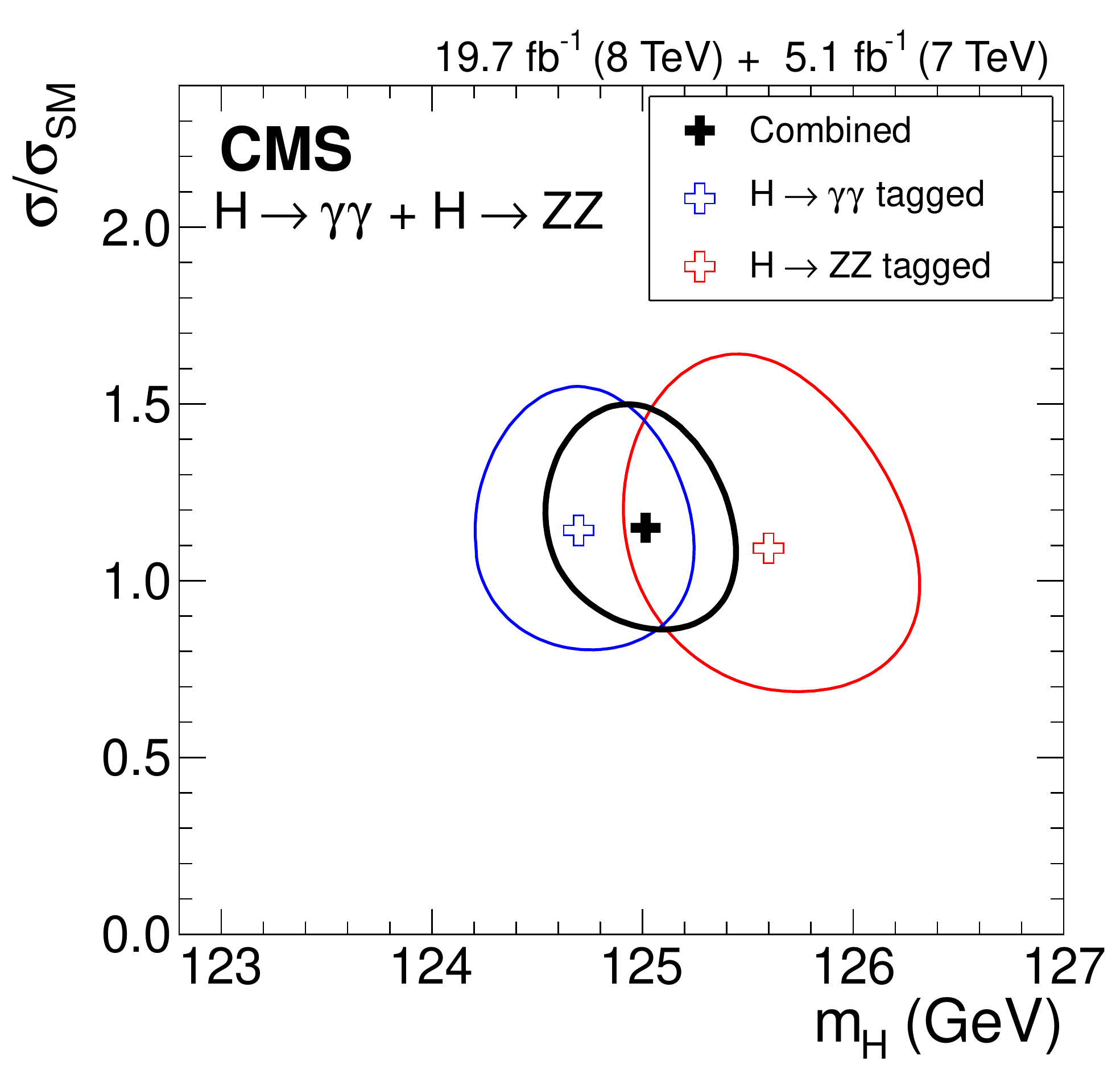}
\caption{\label{mass_c1}The $m_H$ measurement versus the signal strength~\cite{mass_cms}, showing two decay channels and the combined result.} 
\end{minipage}\hspace{0.04\textwidth}%
\begin{minipage}{0.43\textwidth}
\includegraphics[height=6.0cm]{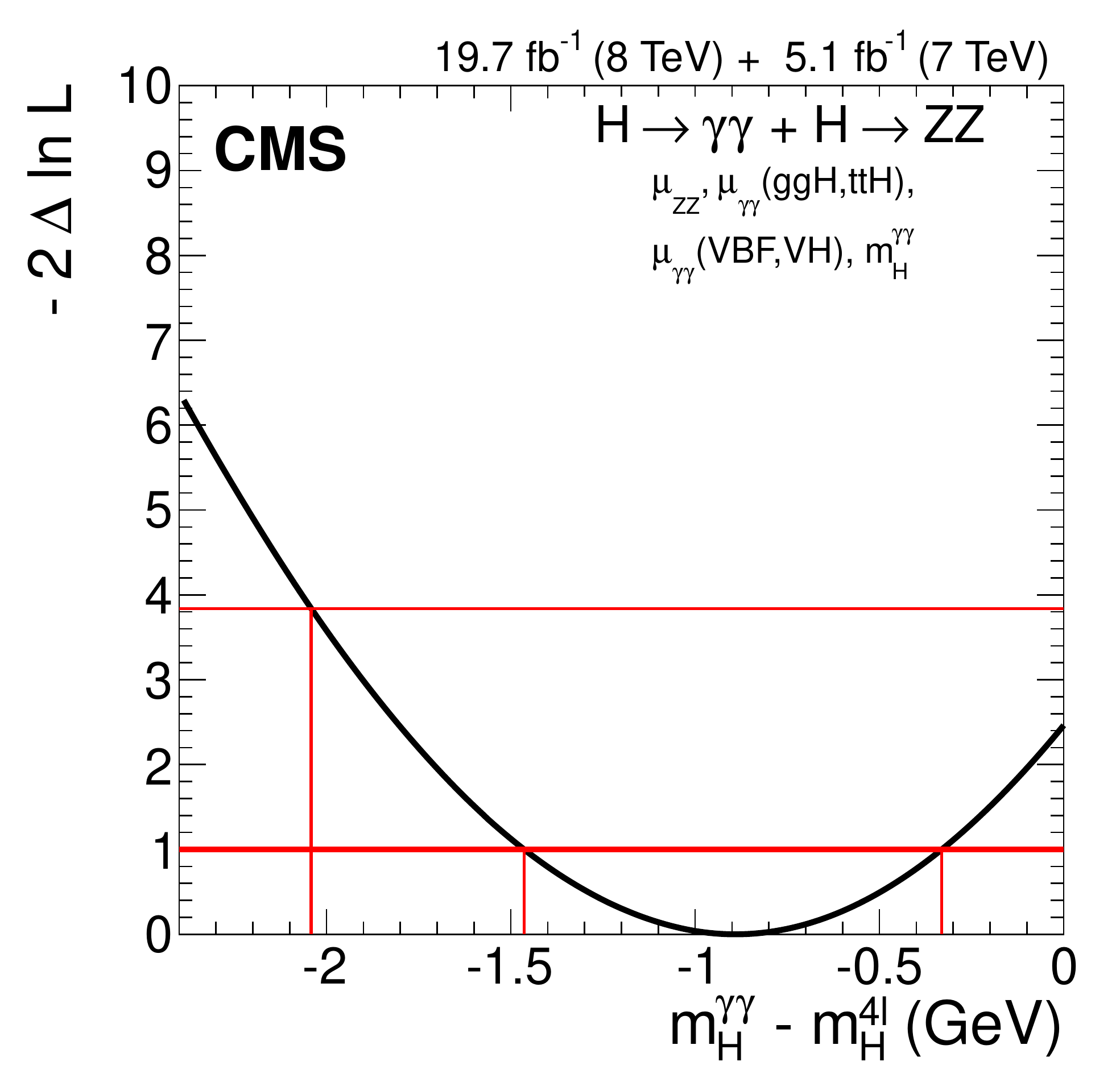}
\caption{\label{massdiff_c1}Likelihood distribution for the difference of the $m_H$ values measured in the $4l$ and the $\gamma\gamma$ final state~\cite{mass_cms}.}
\end{minipage}
\end{center}
\end{figure}
The Higgs boson mass measurement uses input from the $\gamma \gamma$ and 4-lepton decay modes. 
In both channels, the Higgs boson candidate mass can be reconstructed with high precision directly 
from its visible decay products. 
The measurement~\cite{mass_cms} yields $m_H=125.03 ^{+0.26}_{-0.27} \mathrm{(stat)} ^{+0.13}_{-0.15} \mathrm{(syst)}$ 
and is only weakly correlated to the signal strength, see Fig.~\ref{mass_c1}.
The measured four-lepton mass is slightly smaller than the $\gamma\gamma$ mass. 
The tension between the two measurements is about $1.4\sigma$, see Fig.~\ref{massdiff_c1}.
The combination with the corresponding ATLAS measurement leads to the result $m_H=125.09 \pm 0.21 \mathrm{(stat)} \pm 0.11 \mathrm{(syst)}$~\cite{mass_lhc}.

\subsection{Width}
\begin{figure}[htb]
\begin{center}
\begin{minipage}{0.47\textwidth}
\includegraphics[height=7cm]{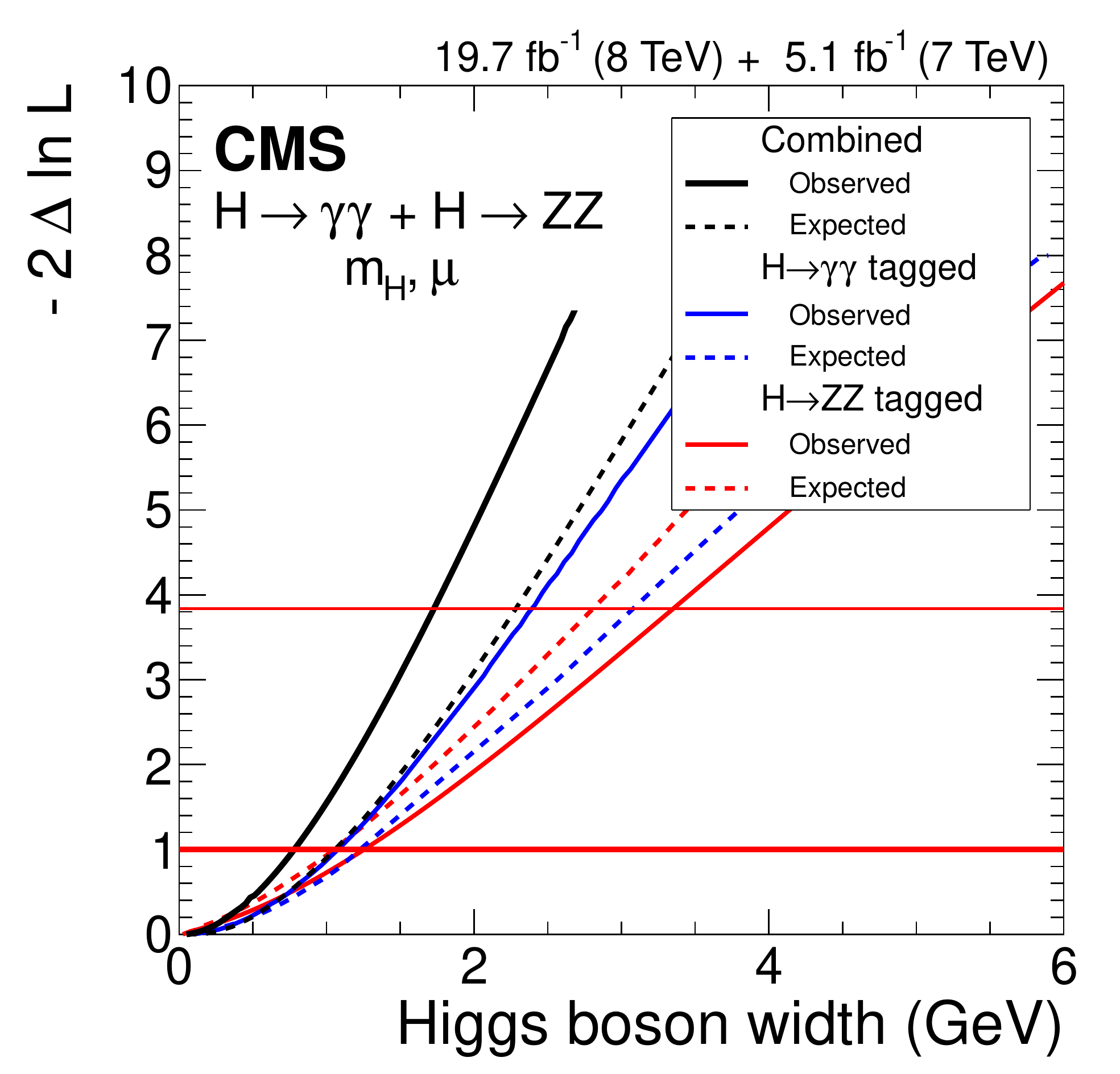}
\caption{\label{width_dir}The direct Higgs boson width measurement in the $4l$ and $\gamma\gamma$ final states~\cite{mass_cms}.}
\end{minipage}\hspace{0.04\textwidth}%
\begin{minipage}{0.47\textwidth}
\includegraphics[height=7.0cm]{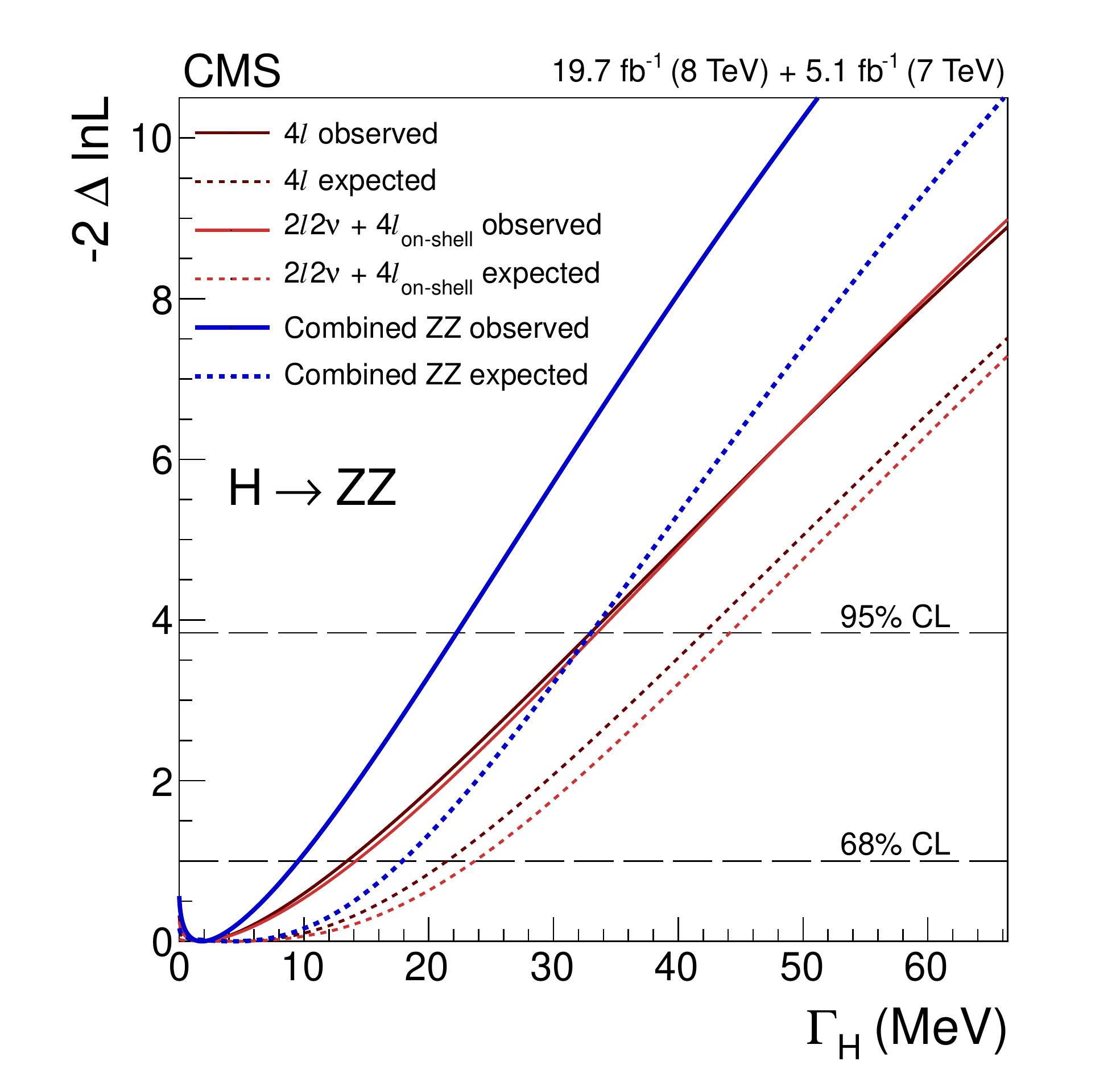}
\caption{\label{width_offsh}The Higgs boson width measurement via off-shell $H \to 4$ leptons signal strength~\cite{width_cms}.} 
\end{minipage}
\end{center}
\end{figure}
The SM expectation for the Higgs boson width at $m_H=125$~GeV is about 4~MeV. There are several ways to access the Higgs boson width 
experimentally. Directly, the width can be measured by analyzing the width of the $m_\mathrm{4l}$ and $m_{\gamma\gamma}$ distributions. 
This method is limited by the experimental resolution which is about three orders of magnitude higher than the width predicted by the SM. 
The direct CMS width limit is 1.7 GeV~\cite{mass_cms}, combining results from the $H \to \gamma\gamma$ and $H \to 4$ leptons channels, see 
Fig.~\ref{width_dir}.

Indirect limits are model-dependent and can be obtained by Higgs boson coupling fits (by leaving the invisible width as free fit parameter) or in direct 
searches of Higgs boson decays to invisible particles. However, the most precise indirect measurement is based on the comparison 
of the on- and off-shell $H \to 4$ leptons signal strength. The observed limit on the Higgs boson width using this technique is 
22 MeV (expected: 33 MeV), which corresponds to 5.4 (expected: 8.0) times the SM expectation~\cite{width_cms}, see Fig.~\ref{width_offsh}.

\subsection{Signal strength}
\noindent
\begin{figure}[htb]%
\noindent
\begin{center}
\begin{minipage}{0.47\textwidth}
\noindent%
\includegraphics[width=7cm]{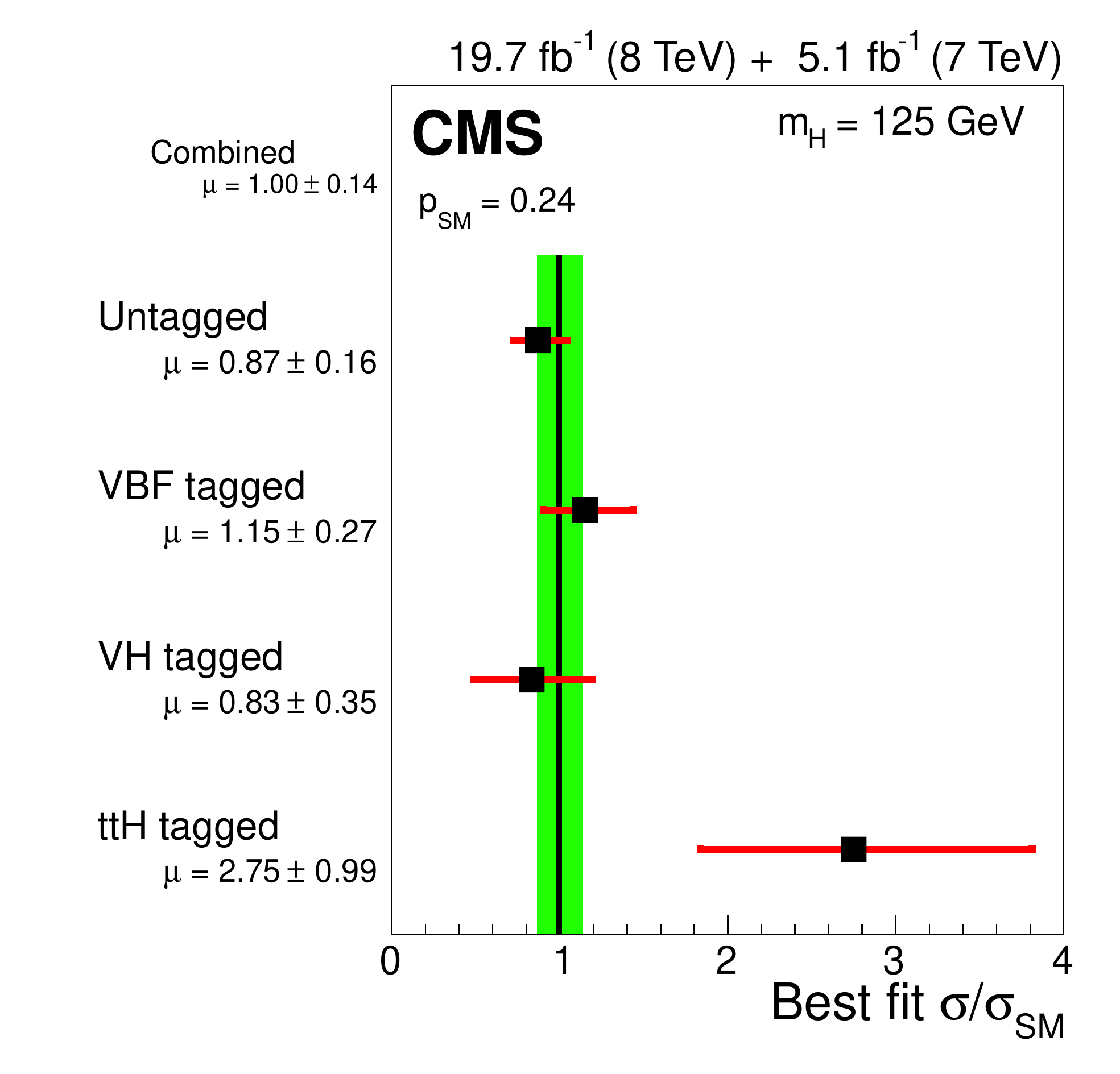}
\caption{\label{mu_c1}Measurement of the production-related signal strength $\mu$~\cite{mass_cms}.}
\end{minipage}%
\hspace{0.04\textwidth}%
\begin{minipage}{0.47\textwidth}
\includegraphics[width=7cm]{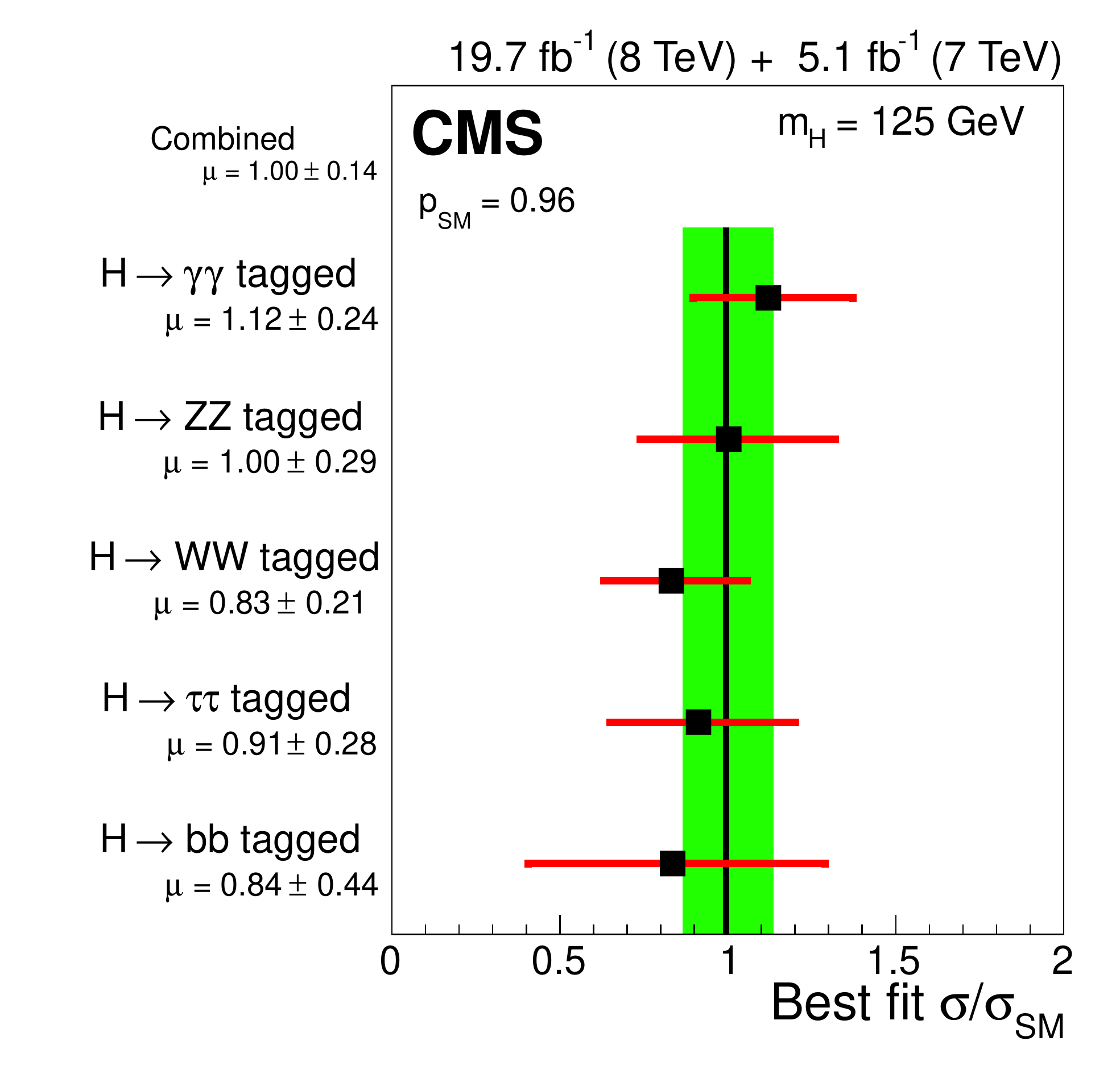}
\caption{\label{mu_a1}Measurement of the decay-related signal strength $\mu$~\cite{mass_cms}.} 
\end{minipage}
\end{center}
\end{figure}
The signal strength $\mu$, defined as measured cross section times branching ratio of a given process divided 
by the SM expectation, is an important test for the validity of the SM. With the data accumulated in the LHC Run 1, 
no significant deviation is observed. The combined result is $\mu = 1.00 \pm 0.13$~\cite{mass_cms}. 
The results split by production mode, shown in Fig.~\ref{mu_c1},  are consistent with the SM in spite of a small excess of events observed
for the $ttH$ mode. Split by decay channel, all measured values are consistent with the SM within $1\sigma$, see Fig.~\ref{mu_a1}. 

\subsection{Coupling strength}
\begin{figure}[htb]
\begin{center}
\begin{minipage}{0.47\textwidth}
\includegraphics[height=6.5cm]{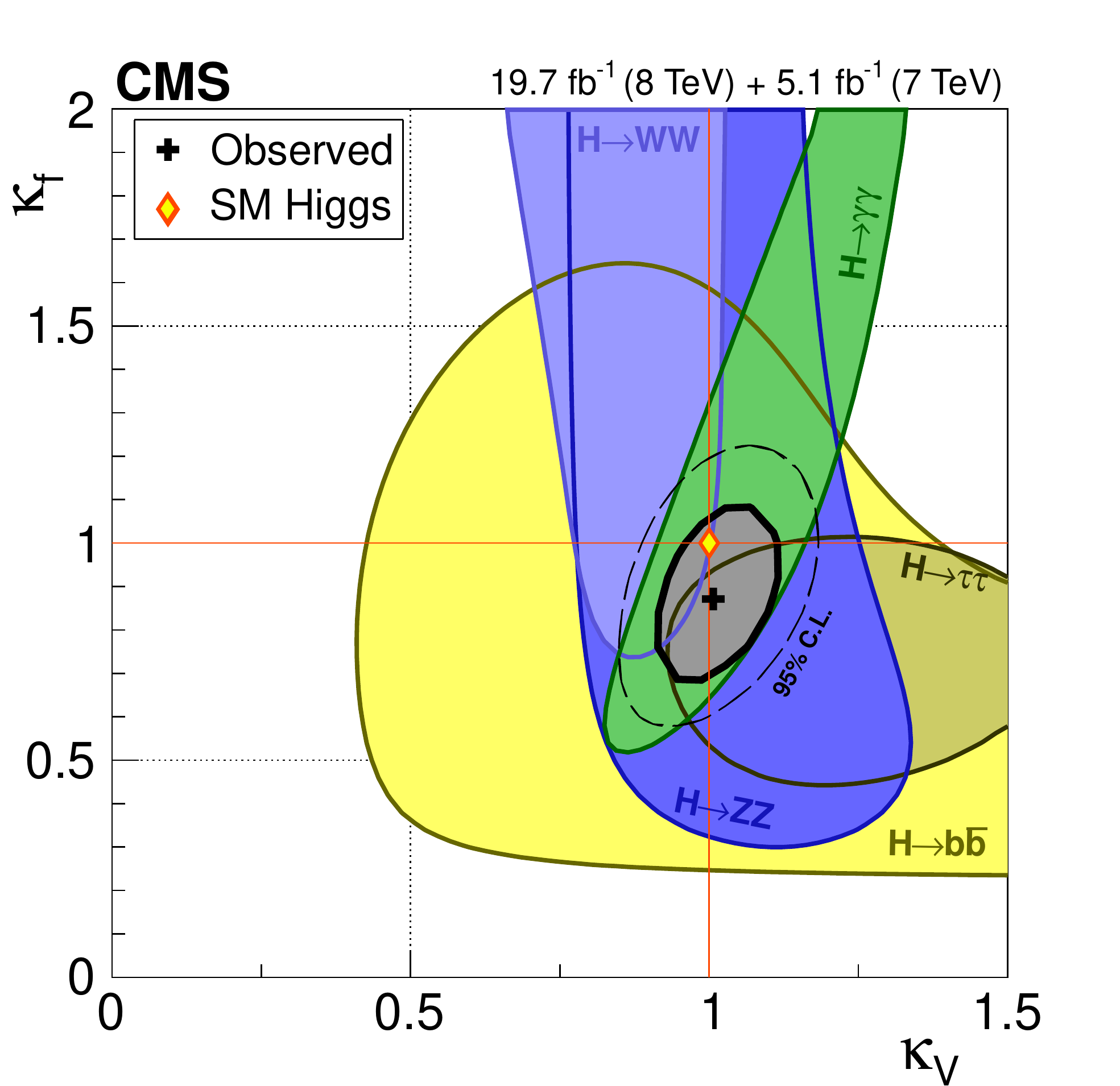}
\caption{\label{kappa_a1}Measurement of $\kappa_V$ versus $\kappa_F$~\cite{mass_cms}.}
\vspace{1\baselineskip}
\end{minipage}\hspace{0.04\textwidth}%
\begin{minipage}{0.47\textwidth}
\includegraphics[height=6.5cm]{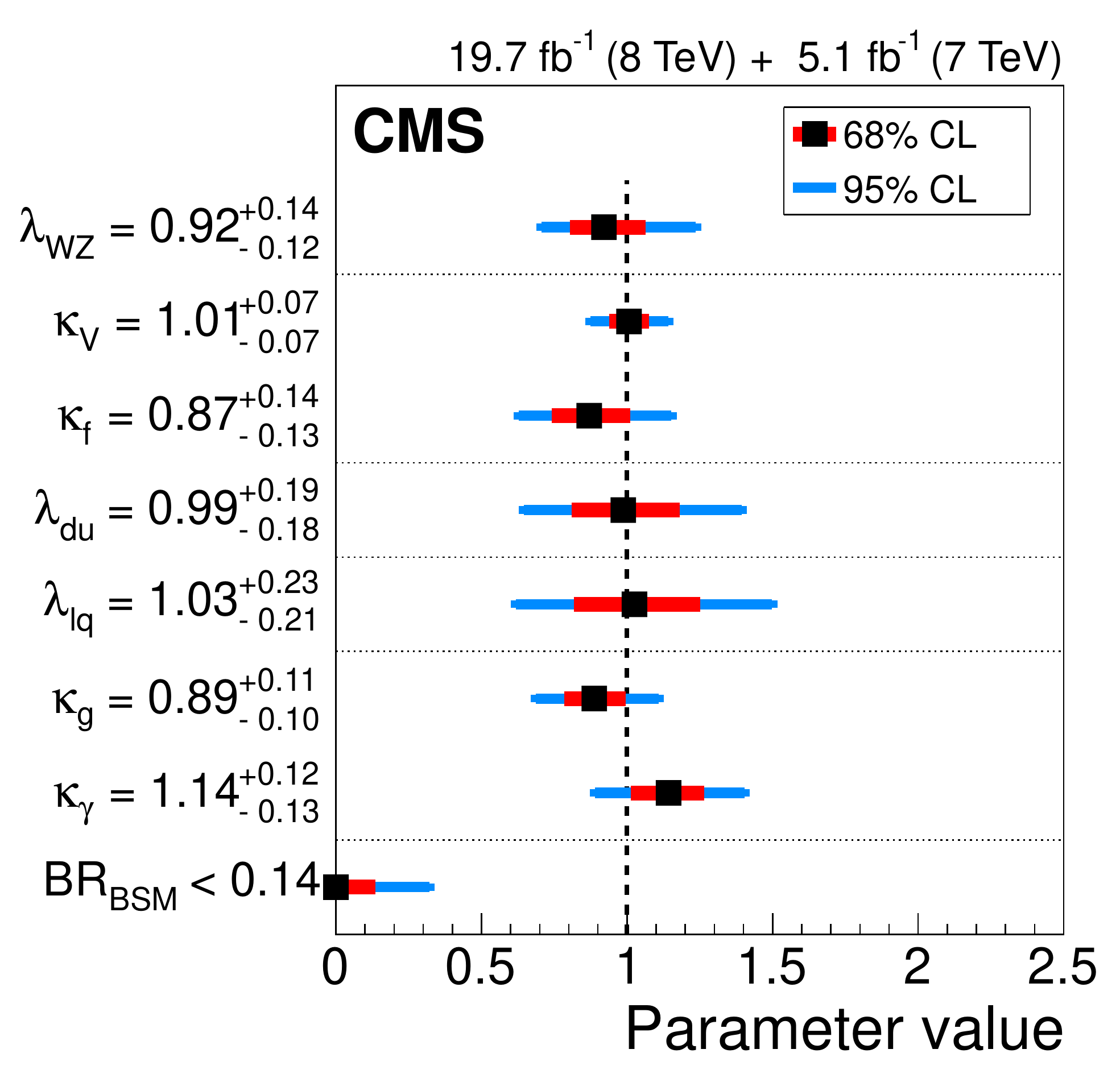}
\caption{\label{kappa_c1}Measurement of coupling strength parameters~\cite{mass_cms}.} 
\end{minipage}
\end{center}
\end{figure}
The coupling strength factors $\kappa_i$ are a leading-order-inspired parametrisation of the Higgs boson coupling to a particle 
or particle class $i$ with respect to the SM expectation. $\lambda_{jk}$ is used to denote the ratio of two values $\kappa_j$ and 
$\kappa_k$. The measurements are typically a result of a global fit of a large subset of the Higgs boson analyses and due to the 
large number of free parameters, some assumptions have to be made to obtain sensible results with the data of the LHC run 1. 
Usual assumptions are those on the Higgs boson width (to be SM-like) or on a universal coupling strength scaling, e.g. of all fermions or all up-type quarks.
An important SM test is the scaling of a fermion coupling, $\kappa_f$, versus a vector boson coupling, $\kappa_V$, shown in Fig.~\ref{kappa_a1}. The result is consistent 
with the SM expectation~\cite{mass_cms}. Further tests involve the custodial symmetry ($W$ versus $Z$ coupling), up- versus down-type quark couplings 
or lepton versus quark couplings. As shown in Fig.~\ref{kappa_c1}, all results agree with the SM prediction~\cite{mass_cms}.

\subsection{Tensor coupling structure}
\begin{figure}[htb]
\begin{center}
\includegraphics[height=4.7cm]{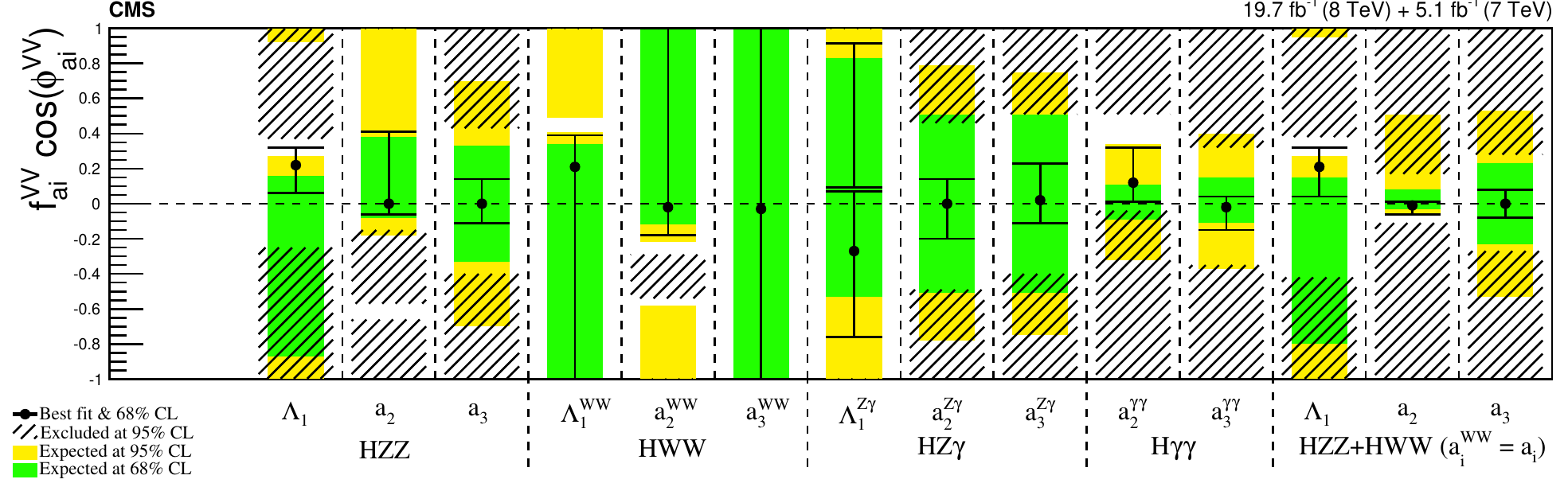}
\caption{\label{cp_c1}Test for anomalous components of the tensor coupling structure of Higgs bosons 
to various other particles~\cite{cp_cms}.}
\end{center}
\end{figure}
The study of the spin and CP properties of the discovered boson are essential for the claim of the discovery of the first fundamental scalar. 
To this date, all measurements are in agreement with the SM but since the parameter space of alternative models is continuous in several 
dimensions, no generic exclusion of e.g. spin-2 models has been possible. However, a large set of the best-motivated models has been tested 
with the result that the SM hypothesis is favored~\cite{cp_cms}. This is illustrated in Fig.~\ref{cp_c1} where measurements of several 
parameters for the spin-0 case are made and all are found to be consistent with the SM expectation. Today, all tested alternatives to a spin-0 boson 
are disfavored and large anomalous contributions to the CP structure of the Higgs boson couplings are excluded. However, small or medium-sized 
CP-odd or anomalous CP-even admixtures are still feasible.

\section{BSM Higgs boson searches}
The programme of BSM Higgs boson searches at CMS is very extensive and cannot be 
presented exhaustively here. In the following, highlights of searches for Higgs bosons 
predicted by the Minimal Supersymmetric extension of the SM (MSSM)
are presented, and other BSM Higgs boson searches are listed and their conclusions are 
summarized.

\subsection{MSSM Higgs boson searches}
\begin{figure}[htb]
\begin{center}
\begin{minipage}{0.47\textwidth}
\includegraphics[height=6.5cm]{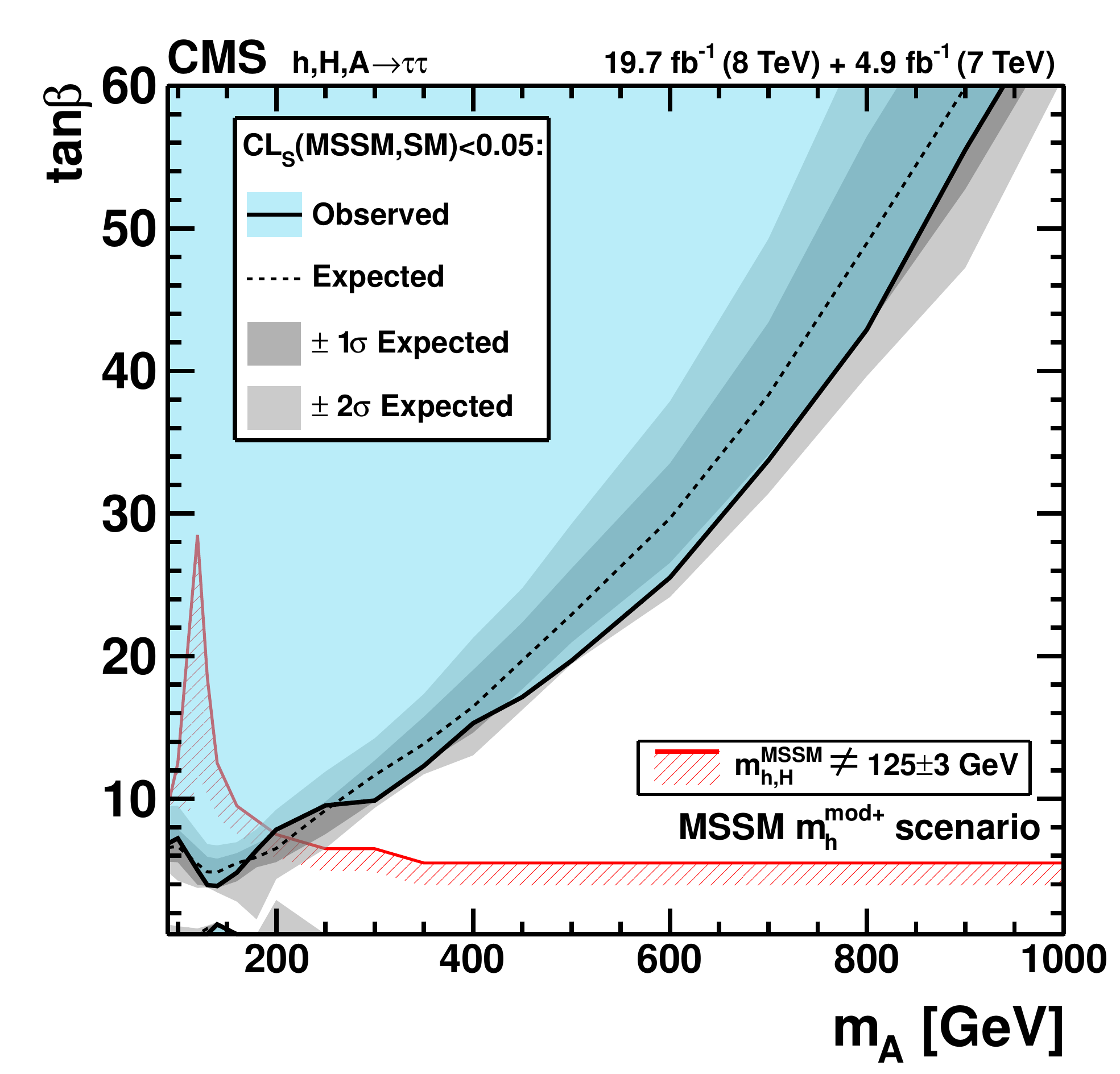}
\caption{\label{mssm_c1}The $m_A$-$\tan\beta$ exclusion contour from neutral MSSM Higgs boson searches~\cite{mssmn_cms}, 
for the $m_h^{\mathrm{mod+}}$ scenario.} 
\end{minipage}\hspace{0.04\textwidth}%
\begin{minipage}{0.47\textwidth}
\includegraphics[height=6.5cm]{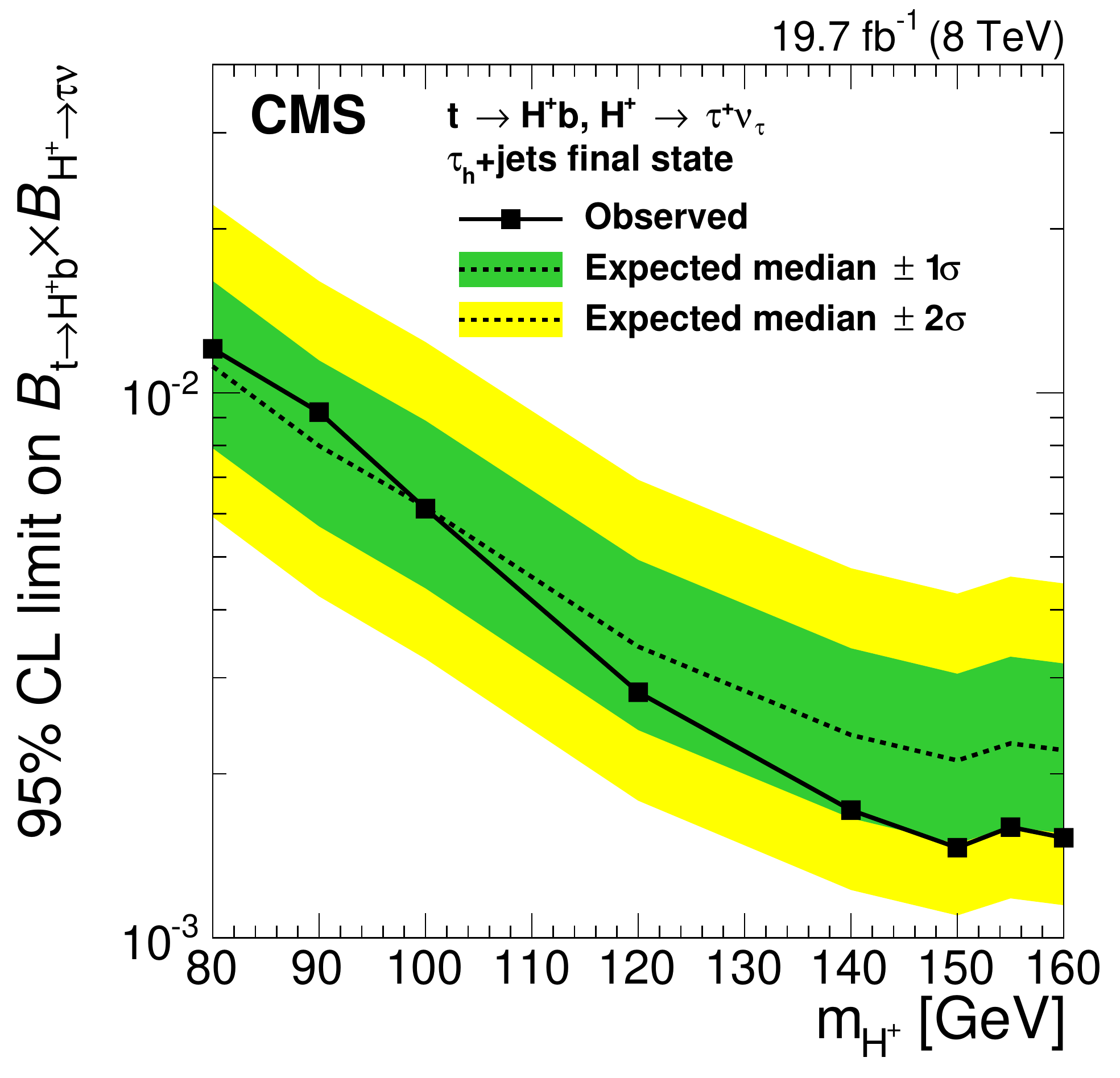}
\caption{\label{mssm_a1}Upper limit on the branching ratio BR($t \to bH^+$)~\cite{mssmc_cms}.}
\vspace{1\baselineskip}
\end{minipage}
\end{center}
\end{figure}
The MSSM predicts the existence of five Higgs bosons, three of them neutral 
and a charged pair. The most sensitive search for a neutral MSSM Higgs boson 
for most of the parameter space is via its $\tau\tau$ decays. The current CMS limit~\cite{mssmn_cms} 
in the $m_h^{\mathrm{mod+}}$ scenario~\cite{mssm_benchmark} is shown in Fig.~\ref{mssm_c1}. 
Due to the charged Higgs boson search result, the option to identify the discovered Higgs boson 
(at $m_H=125$) GeV with the heavy CP-even neutral MSSM Higgs boson has been strongly disfavored. 
Presently, BR($t \to bH^+$) above about 1\% are excluded for $m_{H^+} < 160$ GeV, see Fig.~\ref{mssm_a1}. For charged Higgs bosons 
heavier than the top quark, a sizable region at moderate and high $\tan\beta$ has been excluded~\cite{mssmc_cms}.

\subsection{Other BSM Higgs boson searches}
There is a vast number of BSM Higgs boson searches with CMS in addition to the $A/H \to \tau\tau$ and $H^+ \to \tau\nu$ 
searches introduced in the previous section. A non-exhaustive list of these searches is given here:
%
\begin{itemize}

\item Other (N)MSSM-inspired searches
\begin{itemize}
\item $H^+ \to c\bar{s}$~\cite{cs_cms}
\item $H^+ \to t\bar{b}$~\cite{mssmc_cms}
\item $H/A \to \mu\mu$~\cite{mssm_mumu_cms}
\item $H/A \to b\bar{b}$~\cite{mssm_bb_cms}
\item $a_1 \to \mu\mu$~\cite{amm_cms}
\end{itemize}

\item Generic Higgs boson searches
\begin{itemize}
\item Heavy Higgs, $H \to WW / ZZ$~\cite{heavy_cms}
\item Invisible Higgs, VBF and $ZH$~\cite{inv_cms}
\item Doubly charged Higgs, $H^{++}$~\cite{hpp_cms}
\item Lepton flavor violation, $H \to \tau\mu$~\cite{lfv_cms}
\item Flavor-changing neutral current, $t \to cH$~\cite{fcnc_cms}
\item Fermiophobic Higgs~\cite{fp_cms}
\item Higgs in 4$^\mathrm{th}$-generation models~\cite{fp_cms}
\end{itemize}

\item Indirect search via Higgs boson property measurements~\cite{mass_cms}

\item Higgs-to-Higgs decays, Higgs pair production
\begin{itemize}
\item $HH$ or $X \to HH$~\cite{pair_cms,pair2_cms}
\item $H \to aa$~\cite{aa_cms}
\item $A \to ZH$~\cite{azh_cms}
\end{itemize}


\end{itemize}
%
All searches have in common that no significant deviation from the SM expectation has been found and 
that the phase space for BSM Higgs bosons has been heavily constrained. However, the higher kinematic 
reach at $\sqrt{s}=13-14$ TeV and the extension of the current data set by one or two orders of magnitude 
allows to explore a magnitude of still feasible scenarios over the next few years.

\section{Prospects for Higgs boson searches}
The present LHC programme projects the delivery of about 300 fb$^{-1}$ of data at $\sqrt{s}=13-14$ TeV 
per experiment by the end of 2023. One of the options is that this will be followed by a high-luminosity 
LHC (HL-LHC), producing 3000 fb$^{-1}$ of data at $\sqrt{s}=14$ TeV until 2037. In the following, projections 
of Higgs boson property measurements assuming these conditions are presented. Typically, it is assumed that the 
detector and reconstruction performance is comparable to that of LHC Run 1 due to improved reconstruction 
algorithms and detector upgrades countering the effects of an increase in concurrent proton-proton 
interactions (``pile-up''). Systematic uncertainties are assumed to be similar to the current 
status (Scenario~1) or to go down (Scenario~2; by a factor 0.5 for theoretical uncertainties, 
and by the inverse of the square root of the integrated luminosity for other systematic uncertainties)~\cite{prosp_cms}.
\subsection{Signal strength and coupling strength projections}
\begin{figure}[htb]
\begin{center}
\noindent
\begin{minipage}{0.48\textwidth}
\includegraphics[height=6.0cm]{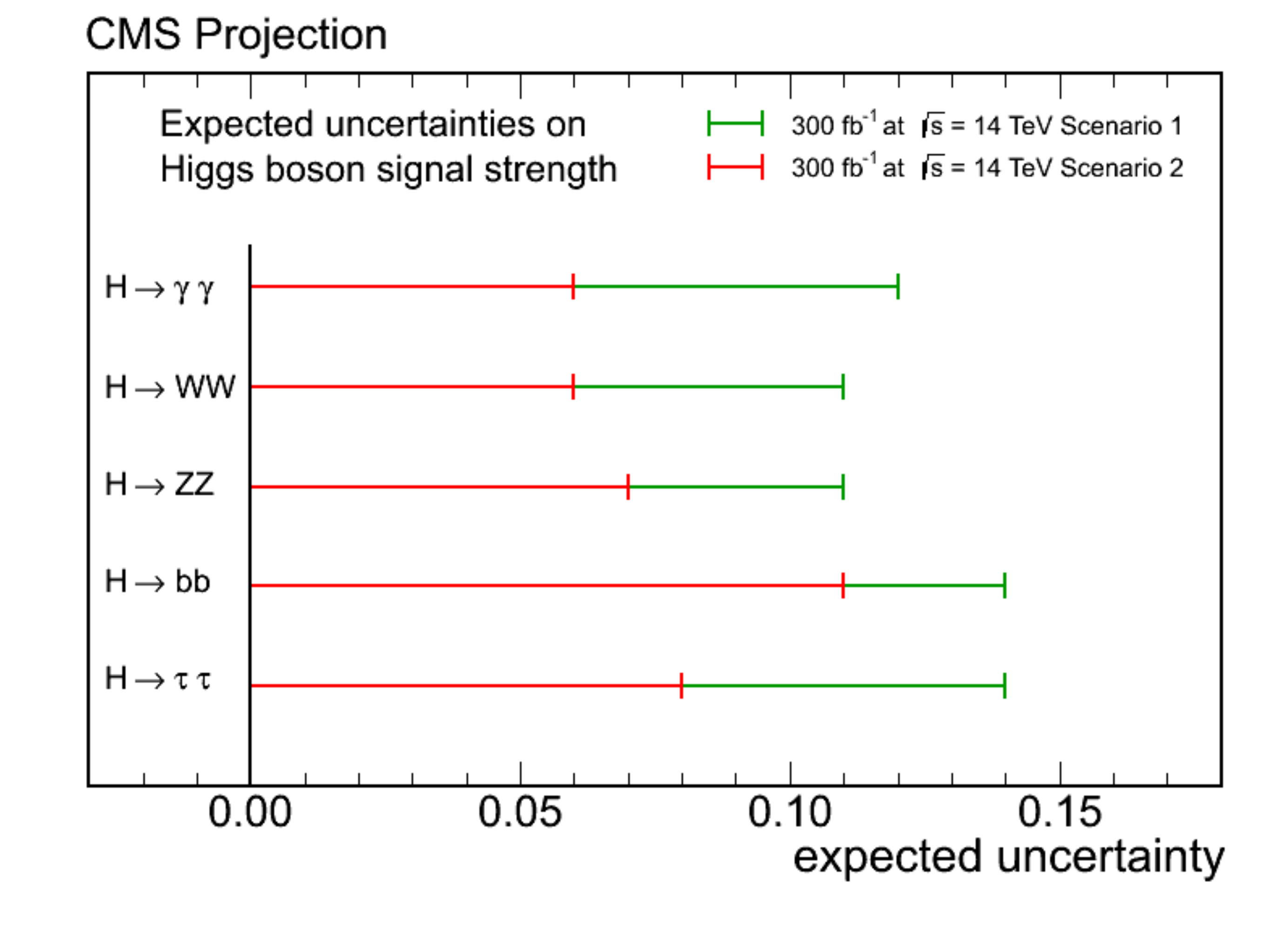}
\caption{\label{prosp_c1}Projection for the precision on Higgs boson signal strength measurements with 300 fb$^{-1}$~\cite{prosp_cms}. The 
scenarios are explained in the text.}
\end{minipage}\hspace{0.04\textwidth}%
\begin{minipage}{0.48\textwidth}
\includegraphics[height=6.0cm]{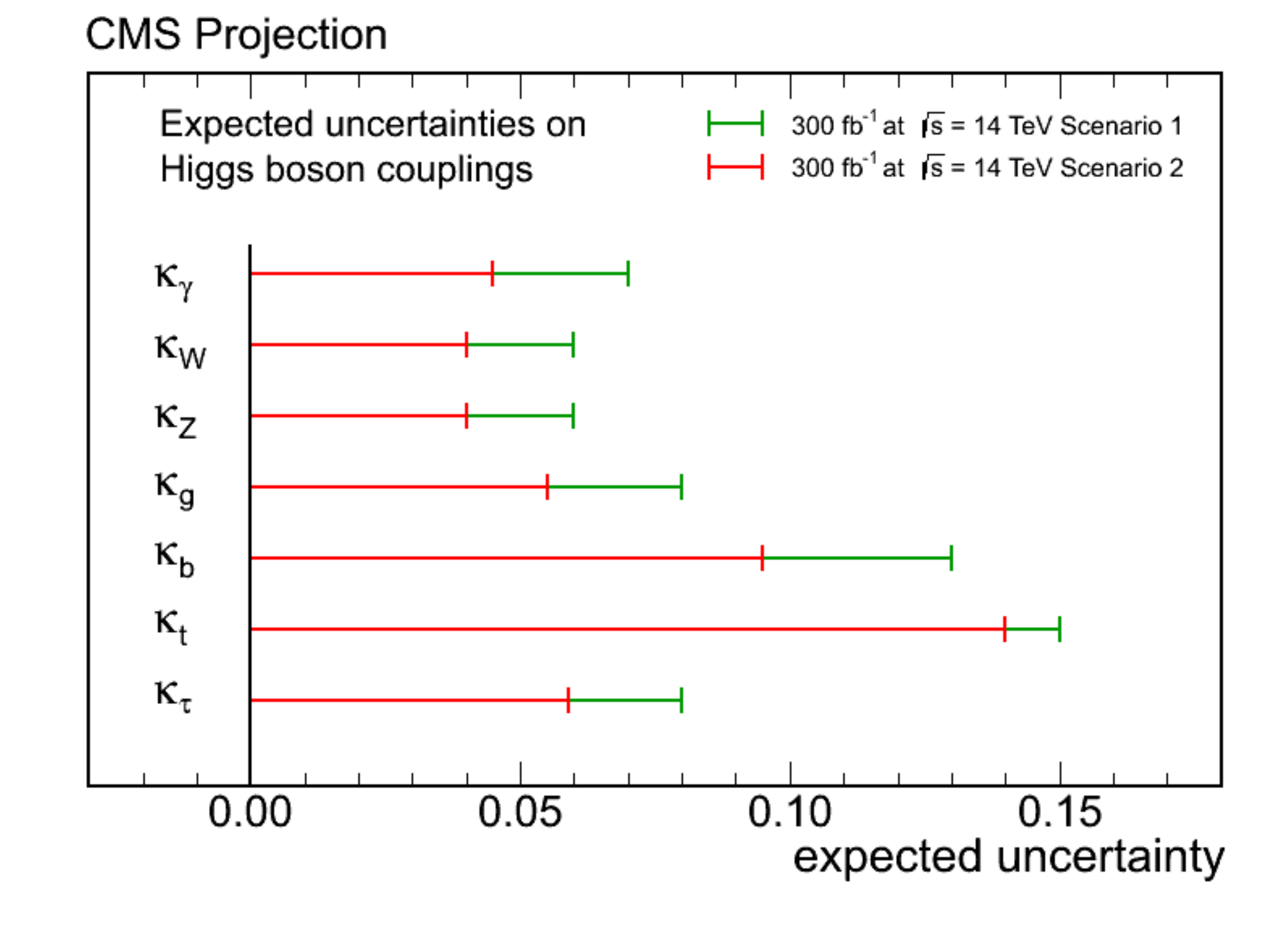}
\caption{\label{prosp_c2}Projection for the precision on the Higgs boson coupling strength measurements with 300 fb$^{-1}$~\cite{prosp_cms}.}
\end{minipage}%
\end{center}
\end{figure}
The estimate for the first 300 fb$^{-1}$ is a precision of roughly 10\% on the signal strength 
for the five most sensitive channels, see Fig.~\ref{prosp_c1}. The 
predicted uncertainties are between 6\%--14\%, with lower values for the boson decay modes 
and higher ones for the fermionic modes, and 
depending on the scenario~\cite{prosp_cms}. This means an improvement of a factor $2-5$ with 
respect to the current measurements. Using 3000 fb$^{-1}$, the precision is expected to go 
down by roughly an additional factor of 2; in addition, the rate of rare channels such as 
$Z\gamma$, $\mu\mu$ and invisible decays can be measured with a precision of $10\%-20\%$.
%

Concerning the coupling strength, the estimated precision is $4\%-8\%$ on the Higgs boson coupling strength to 
elementary bosons and the $\tau$ lepton, and $10\%-15\%$ to the bottom and the top quark 
with 300 fb$^{-1}$~\cite{prosp_cms}, as illustrated in Fig.~\ref{prosp_c2}. 
With 3000 fb$^{-1}$, these uncertainties will be approximately halved. 
This implies an improvement with respect to the current measurements of up to a factor of 10.

\subsection{Double-Higgs boson production and Higgs boson self-coupling projections}
There are two kinds of contributions to double-Higgs boson production: With, and without a triple-Higgs boson vertex. 
A measurement of the cross section of this process thus allows to draw conclusions about the Higgs boson self-coupling. 
Due to the negative interference of these two contributions in the SM, only about 10 signal events are expected 
in the final state $HH \to bb\gamma\gamma$ in 3000 fb$^{-1}$ of data. 
A relative uncertainty on the $HH$ cross section measurement of about 67\%~\cite{tp} is estimated for this final state.
As of today it is thus unclear if any meaningful measurement of the Higgs boson self-coupling is possible.

\section{Conclusions}
CMS offers a rich Higgs physics programme. Following the Higgs boson discovery in 2012, recent years have seen a focus on 
measuring the properties of this Higgs boson. All measurements so far are consistent with the SM expectation but only the sizable improvements 
on their precision expected in the near future will allow to distinguish between the SM and many of its alternatives. In searches for BSM 
Higgs bosons, no evidence for a new state has been found, severely constraining a wide range of BSM models.

\bibliographystyle{iopart-num}
\bibliography{qfthep}

\smallskip

\end{document}